\documentclass[%
 reprint,
 amsmath,amssymb,
 aps,
prd,
]{revtex4-2}

\usepackage{graphicx}
\usepackage{dcolumn}
\usepackage{bm}

\newcommand{\currentfigwid}{1.0}

\usepackage{xspace}
\usepackage{multirow}
\newcommand{\atomF}{$^{19}{\rm F}$\xspace}

\newcommand{\riNa}{$^{22}{\rm Na}$\xspace}
\newcommand{\riArTS}{$^{37}{\rm Ar}$\xspace}
\newcommand{\riArTN}{$^{39}{\rm Ar}$\xspace}
\newcommand{\riBa}{$^{133}{\rm Ba}$\xspace}
\newcommand{\riCs}{$^{137}{\rm Cs}$\xspace}
\newcommand{\riAm}{$^{241}{\rm Am}$\xspace}
\newcommand{\riCf}{$^{252}{\rm Cf}$\xspace}
\newcommand{\riKrEF}{$^{85}{\rm Kr}$\xspace}

\newcommand{\eV}{{\rm eV}}
\newcommand{\keV}{{\rm keV}}
\newcommand{\MeV}{{\rm MeV}}

\newcommand{\aray}{$\alpha$-ray\xspace}
\newcommand{\bray}{$\beta$-ray\xspace}

\newcommand{\gray}{$\gamma$-ray\xspace}

\newcommand{\pe}{{\rm p.e.}}
\newcommand{\cm}{{\rm cm}}
\newcommand{\mm}{{\rm mm}}
\newcommand{\um}{{\rm \mu m}}
\newcommand{\nm}{{\rm nm}}
\newcommand{\ms}{{\rm ms}}
\newcommand{\us}{{\rm \mu s}}
\newcommand{\ns}{{\rm ns}}
\newcommand{\Pa}{{\rm Pa}}
\newcommand{\Bq}{{\rm Bq}}
\newcommand{\MBq}{{\rm MBq}}

\begin{document}

\preprint{APS/123-QED}

\title{Liquid argon scintillation response to electronic recoils between $2.8$--$1275~{\rm keV}$ in a high light yield single-phase detector}

\author{M.Kimura}
 \email{masato@kylab.sci.waseda.ac.jp}

\author{K.Aoyama}

\author{M.Tanaka}

\author{K.Yorita}
 \email{kohei.yorita@waseda.jp}
\affiliation{%
Waseda University, 3-4-1, Okubo, Shinjuku, Tokyo, 169-8555, Japan
}%

\date{\today}

\begin{abstract}
We measure the liquid argon scintillation response to electronic recoils in the energy range of $2.82$ to $1274.6~\keV$ at null electric field.
The single-phase detector with a large optical coverage used in this measurement yields $12.8 \pm 0.3 ~ (11.2 \pm 0.3)~{\rm photoelectron}/\keV$ for $511.0$-$\keV$ \gray events based on a photomultiplier tube single photoelectron response modeling with a Gaussian plus an additional exponential term (with only a Gaussian term).
It is exposed to a variety of calibration sources such as \riNa and \riAm \gray emitters, and a \riCf fast neutron emitter that induces quasimonoenergetic $\gamma$ rays through a $(n, n'\gamma)$ reaction with \atomF in polytetrafluoroethylene.
In addition, the high light detection efficiency of the detector enables identification of the $2.82$-$\keV$ peak of \riArTS, a cosmogenic isotope in atmospheric argon.
The observed light yield and energy resolution of the detector are obtained by the full-absorption peaks.
We find up to approximately $25\%$ shift in the scintillation yield across the energy range and $3\%$ of the energy resolution for the $511.0$-$\keV$ line.
The Thomas-Imel box model with its constant parameter $\varsigma=0.033 ^{+0.012} _{-0.008}$ is found to explain the result.
For liquid argon, this is the first measurement on the energy-dependent scintillation yield down to a few $\keV$ at null field and provides essential inputs for tuning the argon response model to be used for physics experiments.
\end{abstract}

\maketitle

\section{Introduction}
\label{sec:Introduction}
A liquid argon (LAr) scintillation detector has several features that make it attractive for use in various physics experiments to detect ionization particles:
it has efficient conversion of energy deposition into a scintillation light signal,
powerful discrimination between electronic recoil (ER) and nuclear recoil (NR) events based on its scintillation pulse shape,
and benefits from the fact that large quantities of argon are cheaply available.
One promising application of the detector is to search and identify the NR signal possibly induced by a dark matter candidate, weakly interacting massive particles (WIMPs) \cite{agnes2018darkside, ajaj2019search}.
The typical energy of the signal is in the range of a few keV to several hundreds of keV.
Burdensome backgrounds in this search are ER events caused by $\beta$ rays from diffused isotopes (such as \riArTN and \riKrEF) in LAr and $\gamma$ rays from radioimpurities in detector components.
Predicting the measured signal from these background sources is necessary to estimate its contamination in the signal region of interest.
In this context, characterization of the detector response to ER events is crucial for achieving lower energy threshold, suppressing systematic uncertainty related to background contamination, and hence enhancing physics sensitivity of the search.
Furthermore, recently the searches for new particles, such as bosonic dark matter and axion-like particle, have been actively performed using the ER events by xenon (e.g. \cite{akerib2017first, aprile2019light, Aprile2020Observation}), where its scintillation response is well understood \cite{szydagis2011nest, lenardo2015global, szydagis2018noble}, while the one for argon is not fully established yet. 
Therefore this work is essentially important for physics interpretation to extract physics quantity from observed scintillation signal with LAr.

In the LAr detector, a charged particle interaction excites and ionizes the detector medium, resulting in the formation of self-trapped exciton states, ${\rm Ar}_2^*$, through the collision and recombination processes.
The excimer is formed in either a singlet or a triplet state, both of which decay radiatively with vast different lifetimes of approximately $7~\ns$ and $1.6~\us$, respectively \cite{hitachi1983effect}.
The scintillation light spectra from both radiative decays lie in the vacuum ultraviolet (VUV), peaked at $128~\nm$ \cite{heindl2010the}.
As direct detection of the VUV photon at LAr temperature (around $87~{\rm K}$) is technically challenging, it is often downshifted to the visible region where most cryogenic photosensors exhibit peak sensitivity using a wavelength shifter such as 1,1,4,4-tetraphenyl-1,3-butadiene (TPB) \cite{burton1973fluorescence, porter1970nanosecond}.
The recoiled particle and its energy are inferred from the observed photon signal waveform.

In this work, we measure the LAr scintillation response to ER ranging from $2.82$ to $1274.6~\keV$ using a single-phase detector. 
The measurement is performed with a variety of calibration sources including the $2.82$-$\keV$ line of cosmic-ray induced \riArTS.
Owing to a high light collection efficiency (LCE) of the detector, the low energy \riArTS line in the scintillation signal is identified.
Although these kinds of measurement under finite electric field is important as well, we herein focus on the scintillation response at null electric field.
We present the energy dependence of the scintillation yield, as well as the basic properties of this detector such as the observed light yield and energy resolutions of the full-absorption peaks.
The energy dependence of the scintillation yield down to a few $\keV$ is discussed by comparing a model prediction, which is allowed by the use of the \riArTS source.

\section{Experimental apparatus}
\label{sec:Apparatus}
\begin{figure}[tb]
  \centering
  \includegraphics[width=\currentfigwid\columnwidth]{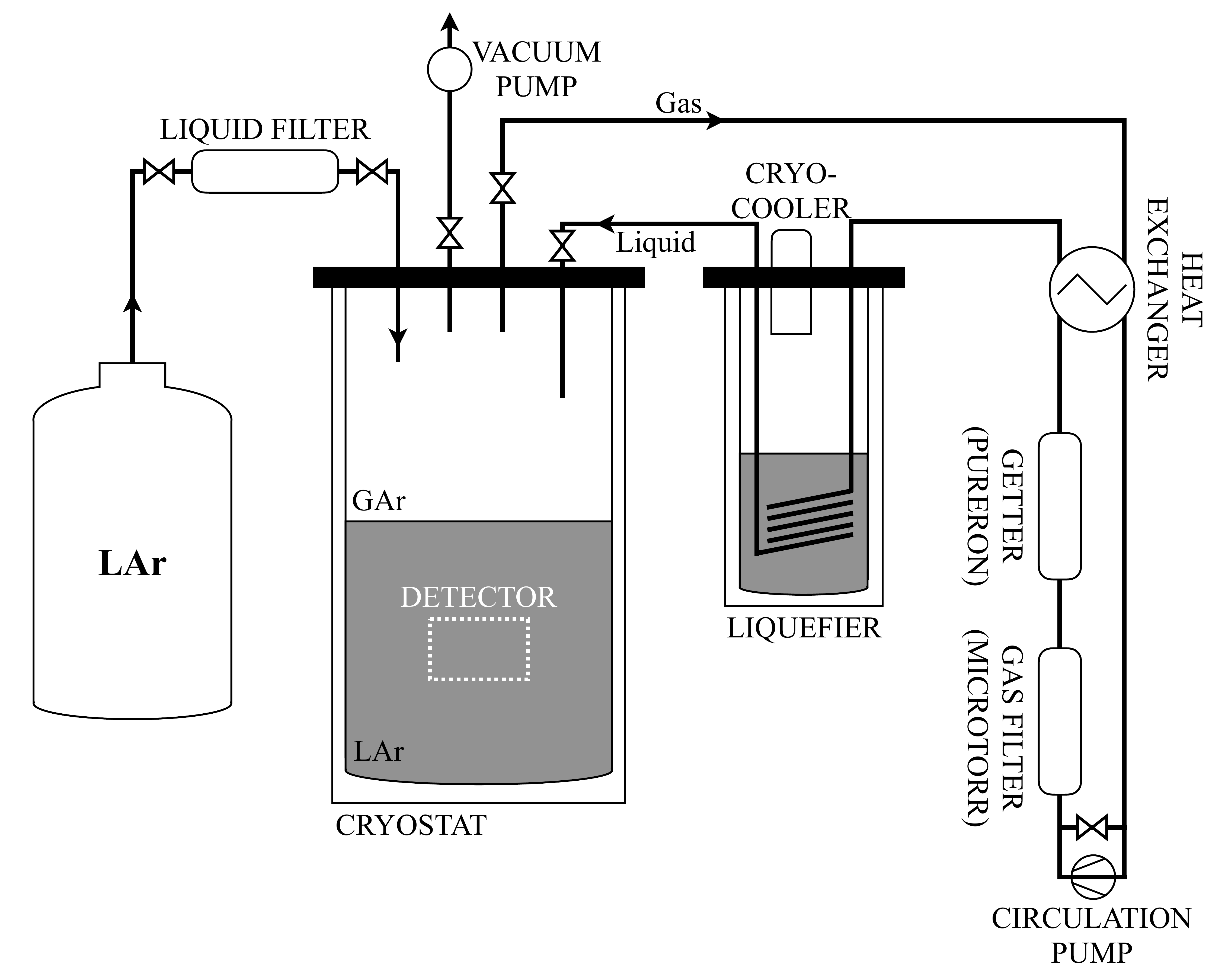}
  \caption{LAr handling system consisting of the filling line (left part of the schematic), the vacuum line (top center), the recirculation line (right), and the main cryostat (center).
  In the recirculation line, gaseous argon (GAr) extracted from the cryostat is pumped into the getters after passing through a heat exchanger.
  It then returns to the heat exchanger to be cooled and is condensed in the liquefier.
  The cryostat containing the detector maintains GAr and LAr over the data collection period in stable cryogenic conditions.}
  \label{fig:WasedaTestStand}
\end{figure}

The measurement presented here is performed at the surface laboratory at Waseda University.
Figure~\ref{fig:WasedaTestStand} shows the argon handling system used in this work.
It mainly consists of a stainless-steel cryostat of diameter $50~\cm$ and height $100~\cm$, in which a scintillation detector sits.
The argon filled in the cryostat is cooled by the recirculation system, which extracts hot gas from the cryostat and passes it through the liquefier with a $200$-${\rm W}$ GM cryocooler (Sumitomo CH-110).
The argon is maintained at a typical pressure of $1.4~{\rm atm}$ and at a liquid level that varies by no more than $1~\mm$ throughout the data collection period.

Impurities in the argon (such as water, oxygen, and nitrogen) affect the scintillation properties, resulting in a reduced signal yield \cite{acciarri2010oxygen, acciarri2010effects, jones2013the}.
In order to remove adsorbed impurities and outgassing from the detector components, the whole system is pumped to vacuum over about ten days before the measurement.
The pressure of the cryostat reaches below $1.0\times10^{-3}~\Pa$.
Then, commercial LAr fills the system via a single path through a liquid filter consisting of a molecular sieve and reduced copper which removes electronegative impurities.
Additional purification is continuously performed by the getters (SAES MicroTorr MC1500-902 and PURERON GP-5) in the recirculation system.
Several measurements performed in this system confirm the concentrations of these impurities are negligible in this measurement:
water and oxygen contaminations of sub-ppb level and nitrogen contamination of sub-ppm level.

\begin{figure}[tb]
  \centering
  \includegraphics[width=\currentfigwid\columnwidth]{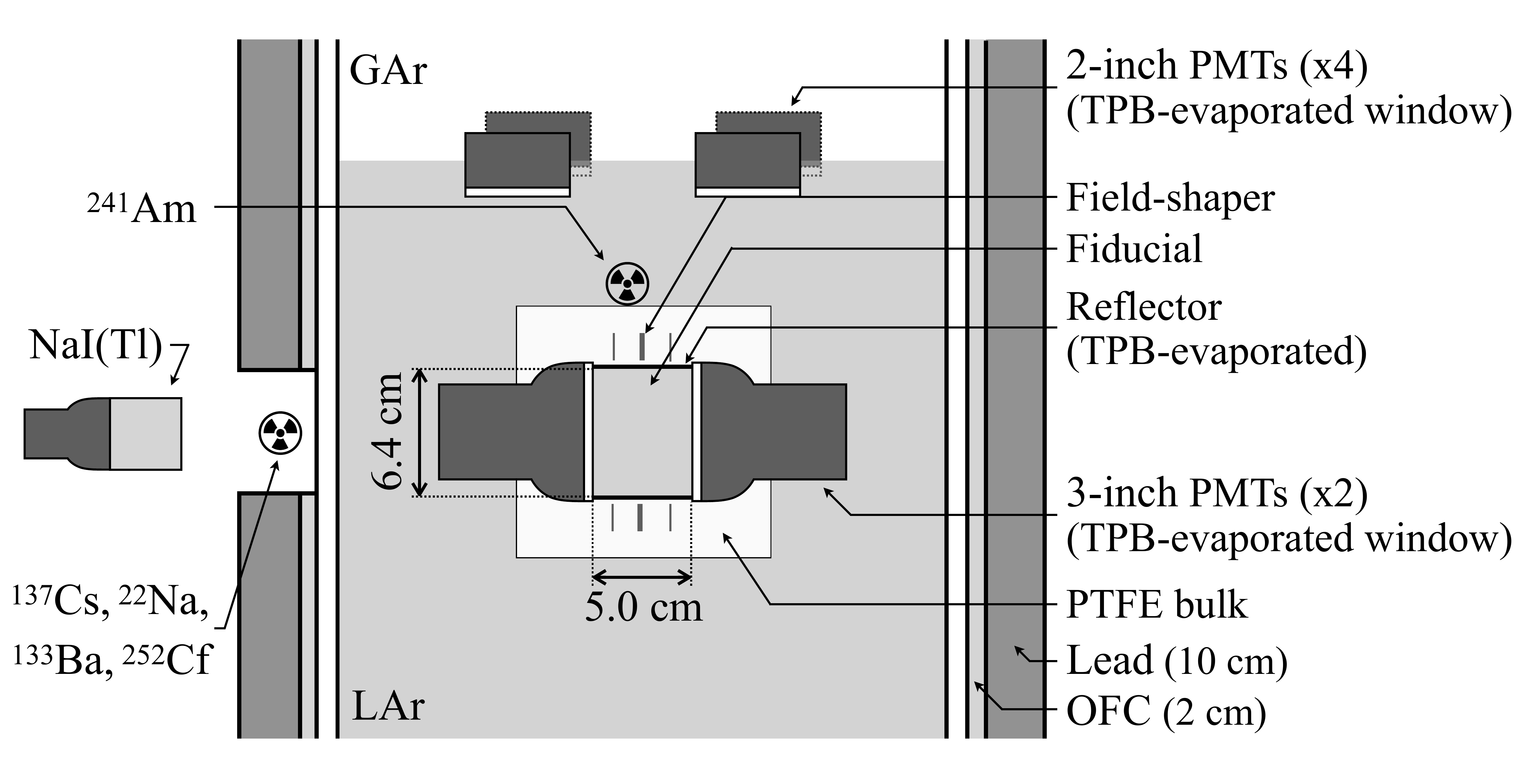}
  \caption{Schematic of the LAr scintillation detector (not scaled).
  The detector including the PMTs is immersed in LAr.
  Oxygen-free copper (OFC) of roughly $2~\cm$ thick and lead of $10~\cm$ thick surround the cryostat and act as a passive shield against ambient $\gamma$ rays.
  An \riAm source is installed at the outer surface of the PTFE bulk, and the other sources (\riCs, \riNa, \riBa, and \riCf) are placed on the outside surface of the cryostat wall.}
  \label{fig:LArDetector}
\end{figure}

The scintillation detector shown in Fig.~\ref{fig:LArDetector} is designed to minimize the loss of scintillation photons in their path and maximize LCE.
The cylindrical fiducial volume of the detector has a diameter $6.4~\cm$ and a length $5~\cm$, contained within an approximately $3$-$\cm$-thick polytetrafluoroethylene (PTFE) sleeve.
The PTFE sleeve serves not only as the main detector structure but also as a \gray emitter, as will be described in Sec.~\ref{subsec:Measurement_252Cf}.
A multilayer plastic-foil reflector (3M ESR) coated with the TPB wavelength shifter lines the inner surface of the PTFE sleeve.
Each end of the cylindrical volume is capped by a $3$-in. Hamamatsu R11065 photomultiplier tubes (PMTs), with around $30\%$ quantum efficiency for blue light after wavelength conversion by the TPB.
The PMT windows are also coated with the TPB.
Both the TPB layer on the reflector and that on the PMT windows are deposited using a vacuum-evaporation technique, and their amounts are approximately $40$ and $30~{\rm \mu g/cm^2}$, respectively, corresponding to the deposited-layer thicknesses of $\mathcal{O}(1~\um)$.
These are confirmed by a quartz crystal microbalance sensor and a stylus profiler, as with a procedure similar to that reported in Ref.~\cite{broerman2017application}.
The $3$-in. PMTs are operated with a negative bias voltage of $-1570~{\rm V}$.
Field-shaping rings with the same bias voltage are embed in the PTFE bulk and ensure electric field inside the fiducial volume less than $1~{\rm V/cm}$ to establish the measurement under null electric field.
The whole sleeve is immersed in a LAr bath contained in the cryostat.

Four $2$-in. PMTs (Hamamatsu R6041-506) are implemented to view the LAr bath surrounding the fiducial volume, as shown in Fig.~\ref{fig:LArDetector}.
These PMTs are located $20~\cm$ above the fiducial volume and just below the liquid surface so that additional energy deposition in the outer region is tagged by a coincident scintillation signal.
The windows of the PMTs are also coated with TPB. 
A passive shield against ambient $\gamma$ rays surrounds the cryostat, which consists of roughly $2$-$\cm$-thick oxygen-free copper and $10$-$\cm$-thick lead.

The data acquisition (DAQ) system used in this experiment consists of a $14$-bit, $250$-${\rm MS/s}$ flash analog-digital-converter (ADC) (Struck SIS3316).
The signals from two fiducial-viewing PMTs and four outer-bath PMTs are digitized and recorded.
The length of the digitizer records is set to $25~\us$ ($5~\us$ before a trigger point and $20~\us$ after), longer than the lifetime of the slow component of LAr scintillation light. 
The trigger is given by the coincidence, within $1~\us$, of the two fiducial PMTs with pulses above a threshold, which is set just above the baseline noise and below a typical single photoelectron ($\pe$) pulse.
The coincidence decision is internally made by the flash ADC board itself.
An inhibition time of $100~\us$ is introduced after each trigger to prevent retriggering of the afterpulse of the PMTs, which mainly occurs after events with far greater energies than the region of interest (e.g., cosmic-ray events).
A Monte Carlo (MC) simulation of the LAr data sample is generated to evaluate the trigger efficiency.
By emulating the internal trigger logic of the flash ADC board on these MC events, the efficiency is found to be consistent with unity for ER signals larger than $25~\pe$, as shown in Fig.~\ref{fig:37Ar_LY}.

\section{Event analysis}
\label{sec:EventAnalysis}
\subsection{PMT calibration}
\label{subsec:Ana_PMTCalib}
\begin{figure}[tb]
	\centering
	\includegraphics[width=\currentfigwid\columnwidth]{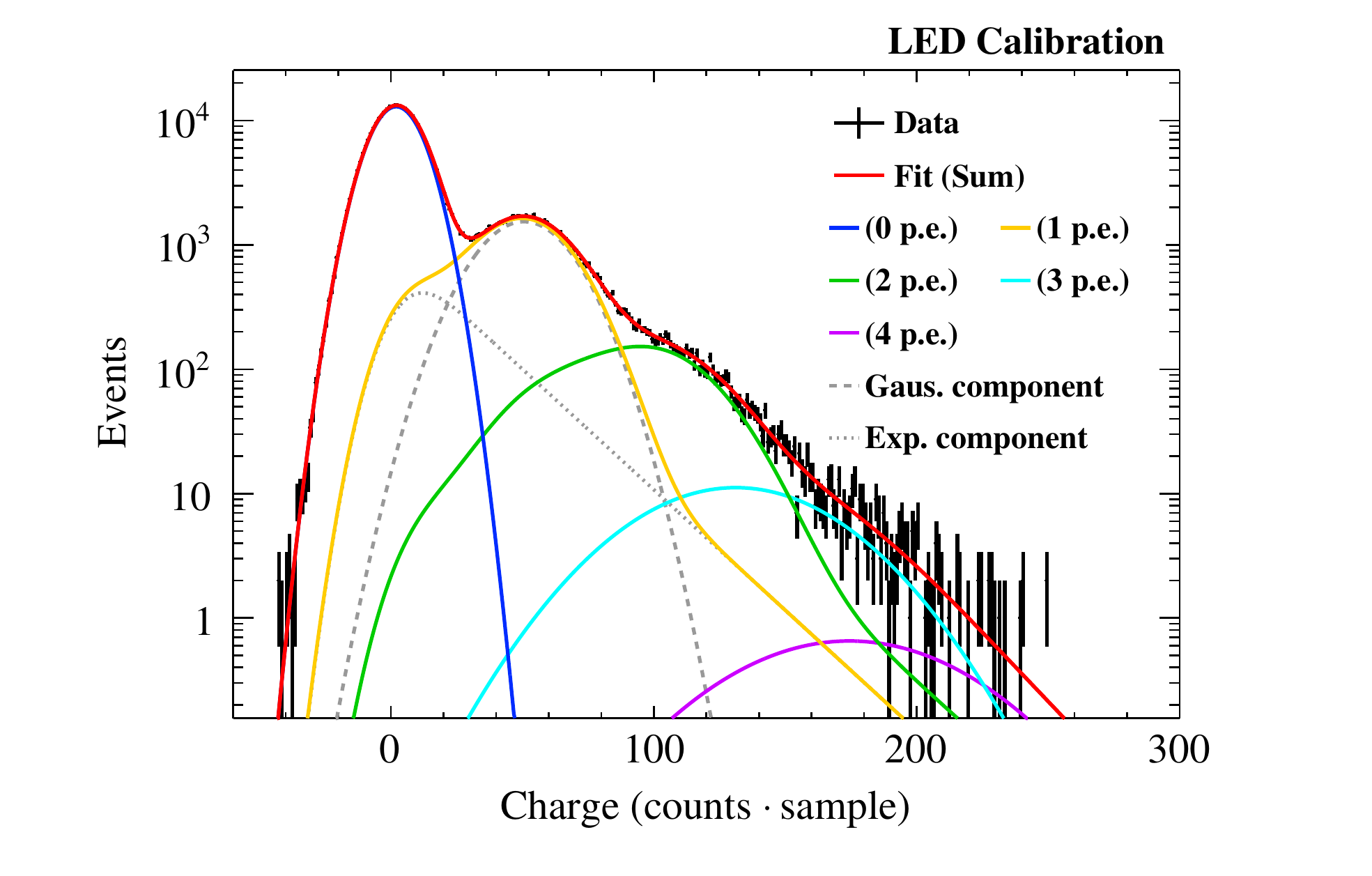}
	\caption{A typical low-light charge distribution of a fiducial-viewing PMT from a LED calibration run.
	The charge is represented in units of integrated digitizer count (${\rm count}\cdot{\rm sample}$), where $1~{\rm count} \cdot {\rm sample}$ corresponds to $9.8 \times 10^{-15}~{\rm C}$.
	The solid red line is the model fit as expressed in Eq.(\ref{eq:GainFit}), and the colored lines represent its components.
	The dashed and dotted lines indicate the Gaussian and exponential terms of the single photoelectron response.}
	\label{fig:LEDCalib_Q}
\end{figure}

The gain of the fiducial-viewing PMTs is calibrated using a blue light-emitting diode (LED) powered by a pulse generator.
Light pulses from the LED characterized by a width of approximately $20~\ns$ at tenth maximum are injected into the fiducial volume through optical fiber, while the generator simultaneously triggers the DAQ system, and the corresponding waveforms from each PMT are recorded over a window of $\pm 1~\us$.
A baseline ADC count is determined by the first $0.6~\us$ of the window, and its subtraction is applied waveform by waveform.
The charge response of the PMT is measured by integrating the waveforms within a $48$-$\ns$ window starting $20~\ns$ prior to the photoelectron pulse arrival time.
The gain value is determined by fitting the charge distribution to model functions.
In this analysis, two models are considered to describe the PMT response.
One expression of the models (gain-model A) as a function of the integrated charge $q$ is followed to that used in Ref.~\cite{Alexander2013light}:
\begin{eqnarray}
	\label{eq:GainFit}
	f(q) &=& \sum_{n} P(n;\lambda) \times f_n (q), \\ \nonumber
	f_n(q) &=& \rho(q) * \psi_1^{n*} (q), \\ \nonumber
	\rho(q) &=& G(q; x_0, \sigma_{\rm ped}), \\ \nonumber 
	\psi_1(q) &=& \frac{p_E}{\tau}\exp(-q/\tau) + (1-p_E) G(q; x_m, \sigma_m) \nonumber
\end{eqnarray}
where 
$P(n;\lambda)$ is a Poisson distribution with mean $\lambda$, 
$G(q; x, \sigma)$ is a Gaussian distribution with mean $x$ and standard division $\sigma$,
$*$ denotes a convolution, 
$\psi_1 (q)$ is the PMT single photoelectron response, 
and $\psi_1^{n*} (q)$ is the $n$-fold convolution of $\psi_1 (q)$ with itself.
This model consists of two components comprising the PMT response:
a simple Gaussian term, which accounts for a photoelectron signal fully amplified by the dynode chain, and an exponential term characterized by a parameter $\tau$, which accounts for underamplified photoelectrons and/or feedback from the dynode photoemission signal.
The fraction of the single photoelectron response found to be the underamplified terms is $p_E$.
Another expression (gain-model B) is simpler, consisting of only the Gaussian term; 
i.e., the fraction $p_E$ in Eq.~(\ref{eq:GainFit}) is fixed to $0$.
This assumes that there is no underamplified or dynode-feedback response in a PMT and that the photoelectron response is perfectly described by Gaussians.

Figure \ref{fig:LEDCalib_Q} shows the charge distribution and fit for a LED calibration run with the gain-model A (which has a nonzero fraction $p_E$), where $1~{\rm count} \cdot {\rm sample}$ corresponds to an output charge of $9.8 \times 10^{-15}~{\rm C}$.
The mean charge for a single photoelectron $g$ defined as
\begin{equation}
	\label{eq:AverageGain}
	g = p_E \tau + (1-p_E) x_m
\end{equation}
is approximately $2.0 \times 10^6~e^{-}/\pe$ with a bias voltage of $-1570~{\rm V}$.
The fit with the gain-model B (i.e., simple convolution of Gaussian functions) returns a $12\%$ higher gain value than gain-model A.
This difference is nearly consistent with the result reported in Ref.~\cite{Alexander2013light}.
While we do not have enough data to determine which model is more appropriate to describe the PMT response, the gain-model A is adopted as  the baseline, and the result from the model is used in the later analysis.
This calibration is performed every $12~{\rm hours}$ during a data collection period lasting seven days.
The overall stabilities of the gain and observed light yield during the period are within less than $0.5\%$ from both the LED measurement and an energy calibration mentioned below.

The nonlinearity of the PMT is studied by a pulsed laser source, and we found that the effect is less than $1\%$ ($0.1\%$) at $1~\MeV$ (below $200~\keV$) at the operation voltage. 
The observed light yields are corrected accordingly, and its correction factors are considered as a systematic uncertainty. 

\subsection{Signal analysis and selection criteria}
\label{subsec:Ana_Analysis}
The analysis of the LAr scintillation signal is performed following a photon-counting algorithm.
For each waveform, this algorithm first calculates the baseline from the pretrigger window;
once that baseline is subtracted, all samples above a software threshold are grouped with three neighboring samples (one bin before and two bins after).
The software threshold is set based on the baseline noise and is below a typical single photoelectron PMT pulse.
The signal detection time is identified as the first sampling time above a threshold of $50\%$ peak amplitude.
Detected scintillation light is defined as the integrated charge in the time interval between $-0.04$ and $7.0~\us$.
A pulse shape discrimination (PSD) parameter is also defined as the fraction of light detected after $0.1~\us$ of the scintillation signal (termed ``slow/total'').

\begin{figure}[tb]
	\centering
	\includegraphics[width=\currentfigwid\columnwidth]{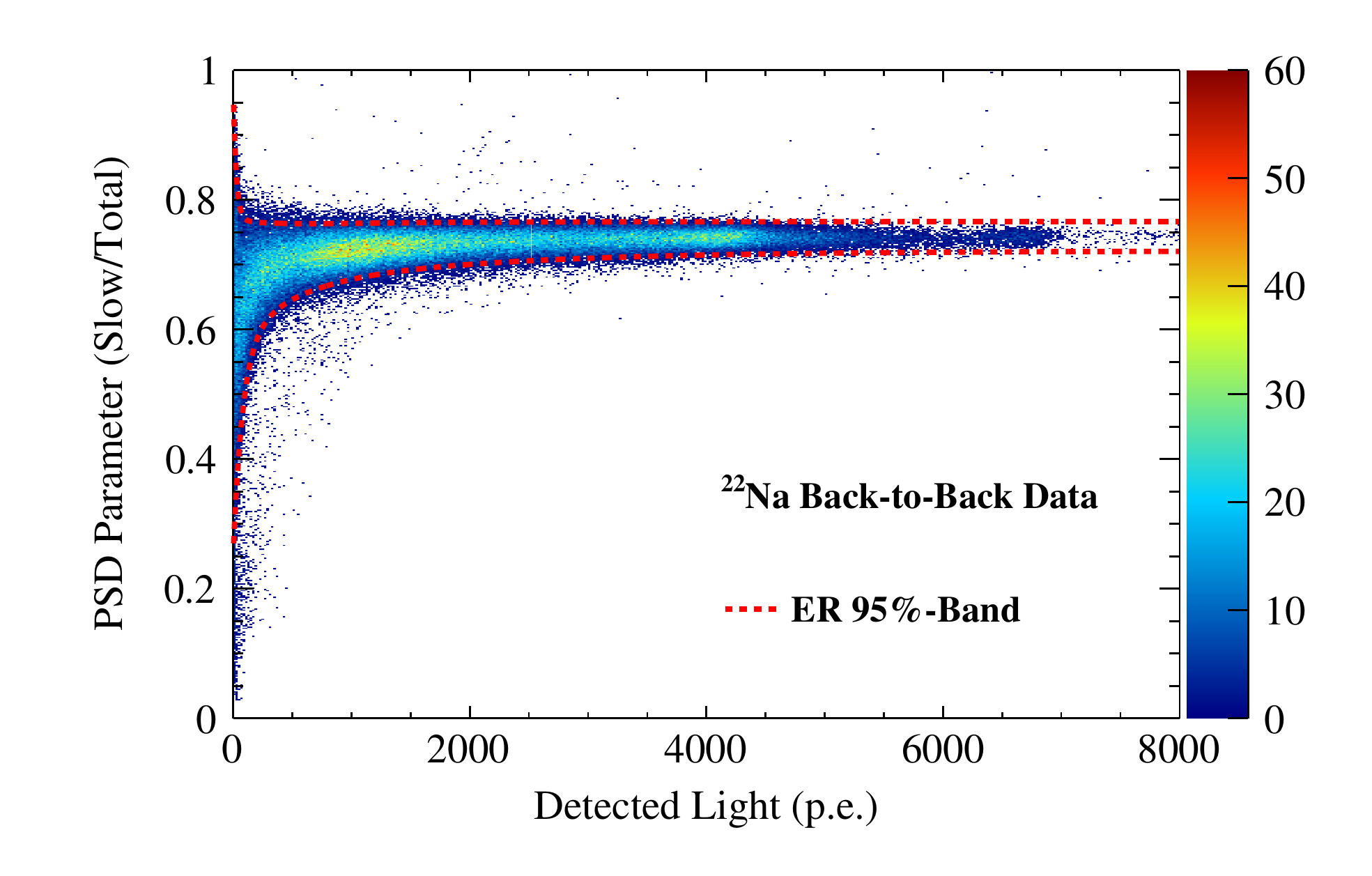}
	\caption{Distribution of the PSD parameter (``Slow/Total'') versus the observed light signal. 
	The data require the back-to-back tagging described in Sec.~\ref{subsec:Ana_ECalib}.
	The red dashed lines correspond to the $95\%$ containing band for ER events.}
	\label{fig:22Na_PSD}
\end{figure}

A set of data quality cuts is applied to remove instrumental effects and event pileups.
The selection criteria are as follows:
(1) Software imposes a $10$-$\ms$ veto after events that contain signals greater than $\approx$$2.0 \times 10^4$ ($\approx$$5.0 \times 10^3$ $\pe$) for datasets taken with a \gray source with $>$$100~\keV$ ($<$$100~\keV$) its energy.
This aims to remove the unstable period of the PMT after outputting a large charge signal.
(2) The event has a stable baseline noise and no more than $0.7~\pe$ pulses in the pretrigger window.
(3) The event does not occur near the PMT and is more likely to be a LAr scintillation signal than Cherenkov light on the PMT window.
The signal asymmetry defined as $A = (N_\pe^1 - N_\pe^2)/(N_\pe^1 + N_\pe^2)$ in which $N_\pe^1$ and $N_\pe^2$ are the observed photoelectron signal in each PMT is used to evaluate the interacting position.
The cut value is selected to contain approximately $99\%$ of the LAr signal.
(4) The PSD parameter of the event is consistent with that of the ER.
This requirement is particularly important for the \riCf data because it enhances the \gray full-absorption peaks over continuous nuclear recoil spectrum.
The band of the parameter used in this cut is determined by \riNa data requiring the coincidence detection of the backward-traveling $511$-$\keV$ $\gamma$ ray whose details are described in the following section.
The selection band contains $95\%$ of ER events, as shown in Fig.~\ref{fig:22Na_PSD}.

\subsection{Determination of photoelectron per keV with sodium-22 and cesium-137 sources}
\label{subsec:Ana_ECalib}
\begin{figure}[tb]
	\centering
	\includegraphics[width=\currentfigwid\columnwidth]{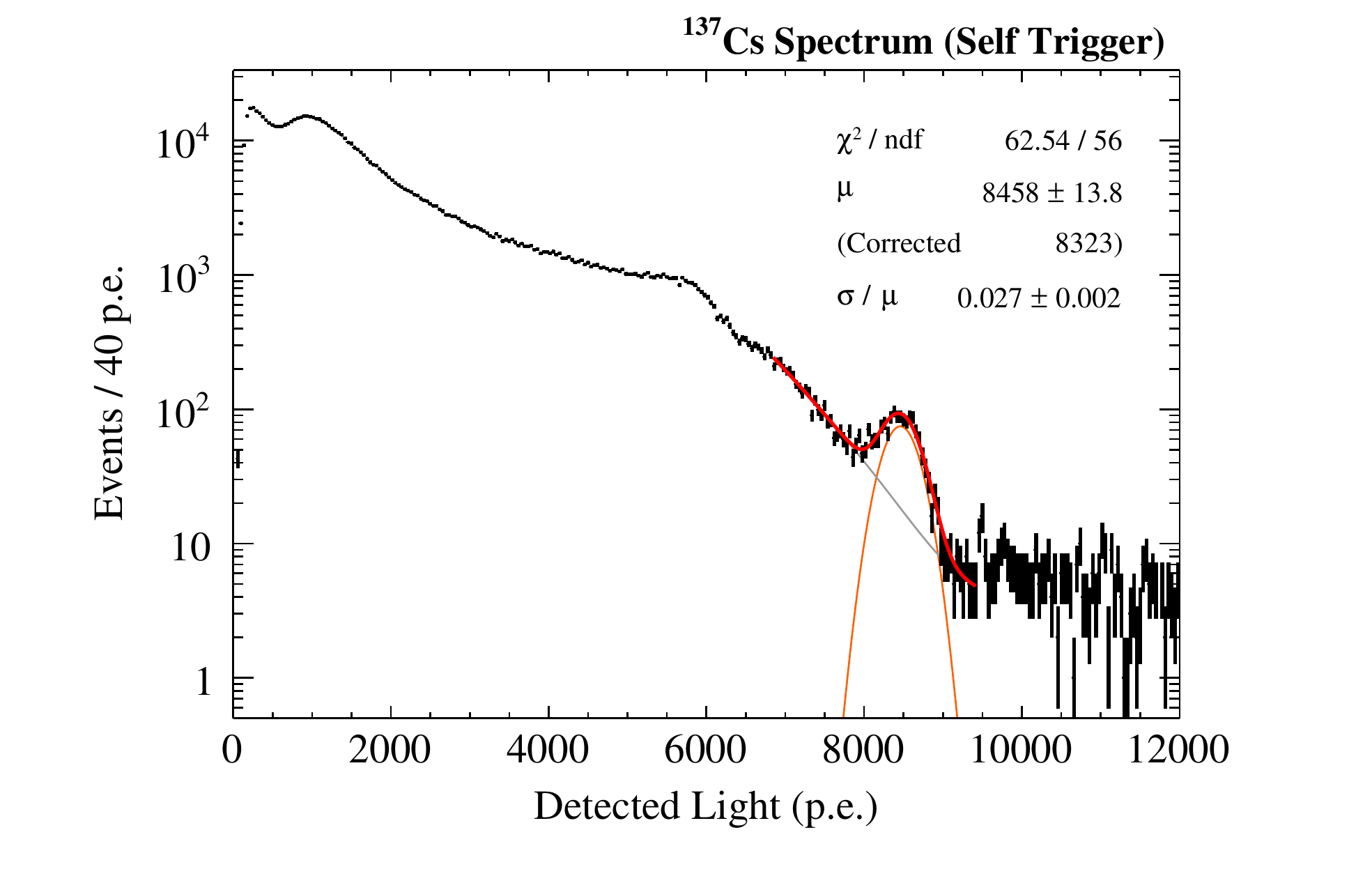}
	\caption{The observed light spectrum from the \riCs source used for the energy calibration.
	The red lines represent the fit function for the $661.7$-$\keV$ peak.}
	\label{fig:137Cs_LY}
\end{figure}
\begin{figure}[tb]
	\centering
	\includegraphics[width=\currentfigwid\columnwidth]{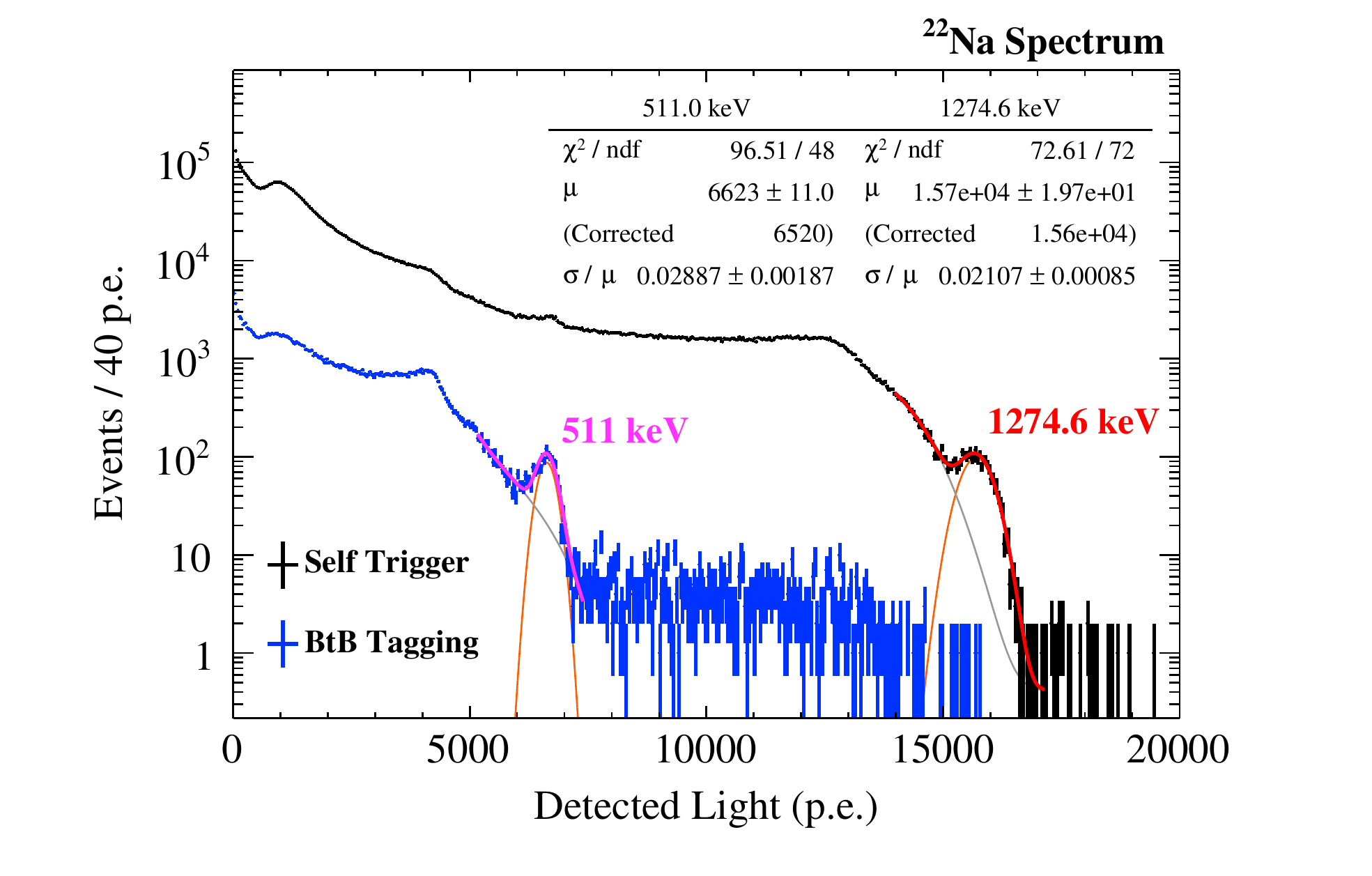}
	\caption{The observed light spectra from the \riNa source before and after requiring back-to-back coincidence (BtB tagging).
	The red and magenta lines represent the fit function for the $1274.6$-$\keV$ peak in self-trigger data and the $511$-$\keV$ peak in back-to-back data, respectively.}
	\label{fig:22Na_LYs}
\end{figure}

\begin{table}[tb]
	\caption{Fitted \gray energy $E_\gamma $ and observed light yields resulting from the full-absorption peak fit.
	The uncertainties listed in the table are combined with both statistical and systematic uncertainties.}
	\begin{tabular*}{\columnwidth}{@{\extracolsep{\fill}}llll} \hline \hline
		\multirow{2}{*}{$E_\gamma ~[\keV]$}	& \multirow{2}{*}{Source}		& \multicolumn{2}{c}{$\mu/E_\gamma~[\pe/\keV]$}	\\\cline{3-4}
						&			& (Gain-model A)			& (Gain-model B) \\ \hline					
		$511.0$			& \riNa		& $12.8 \pm 0.3$ 			& $11.2 \pm 0.3$ \\
		$661.7$			& \riCs		& $12.6 \pm 0.3$ 			& $11.1 \pm 0.3$ \\
		$1274.6$			& \riNa		& $12.3 \pm 0.3$ 			& $10.8 \pm 0.3$ \\ \hline \hline
	\end{tabular*}
	\label{tab:ECalib}
\end{table}

Determination of the observed light yield, photoelectron per keV of the detector is performed by $511.0$-, $661.7$-, and $1274.6$-$\keV$ $\gamma$ rays.
The \gray sources, \riCs and \riNa, with approximately $1~\MBq$, respectively, are placed on the outside surface of the cryostat wall to expose the $\gamma$ ray to the detector.

Figure~\ref{fig:137Cs_LY} shows the observed light spectrum obtained with the \riCs source.
The full-absorption peak of the $661.7$-$\keV$ line of the \riCs source is fit with a Gaussian with mean $\mu$ and width $\sigma$.
The continuous background components around the peak, mainly coming from the Compton edge and degraded tails, are modeled with error and linear functions and added to the fit function.
The fit shown in Fig.~\ref{fig:137Cs_LY} returns $\chi^2/ndf=62.5/56$.

The observed light spectra obtained with the \riNa source are shown in Fig.~\ref{fig:22Na_LYs}.
In this measurement, an additional NaI(Tl) scintillator ($2 \times 2~{\rm in.}^2$ cylinder) is set with the source at opposite sites of the cryostat to tag the backward-traveling $511.0$-$\keV$ $\gamma$ ray (back-to-back tagging).
The distance between the cryostat wall and the source is set to $15~\cm$, and that between the source and the scintillator to $25~\cm$.
The black and blue spectra in Fig.~\ref{fig:22Na_LYs} are the observed scintillation spectra before and after requiring the coincidence detection of the $511.0$-$\keV$ \gray signal in the NaI(Tl) scintillator.
Since the $1274.6$-$\keV$ $\gamma$ ray is considered to have no angular correlation with back-to-back $\gamma$ rays, the corresponding peak appears only in the former spectrum.
Each peak is fit with a Gaussian plus background model function consisting of error and linear functions.
Values of $\chi^2/ndf=72.6/72$ and $\chi^2/ndf=96.5/48$ are returned from the fits for $1274.6$- and $511.0$-$\keV$ peaks, respectively.

These observed photoelectron signals contain extra charge from PMT afterpulses and systematic effect from the photon-counting algorithm.
A correction for these effects is thus applied to reconstruct the observed light signal per ER energy.
This correction is based on an independent study of the PMT response as well as a MC simulation of the LAr signal.
It is relatively small, approximately $1\%$ for the \riCs line and less than $3\%$ for the whole energy region of interest of this analysis, where the amount of afterpulse is estimated as $2\%$--$4\%$ of the photoelectron signal, and the algorithm can systematically slightly underestimate the charge signal.
The observed light yields after the corrections are summarized in Table~\ref{tab:ECalib} with uncertainties.
The uncertainty includes the estimation of PMT afterpulses, systematic error in the corrections, and stability of the detector.

\section{Measurement of scintillation response with calibration sources}
\label{sec:Measurments}
\subsection{Barium-133 source}
\label{subsec:Measurement_133Ba}
\begin{figure}[tb]
	\centering
	\includegraphics[width=\currentfigwid\columnwidth]{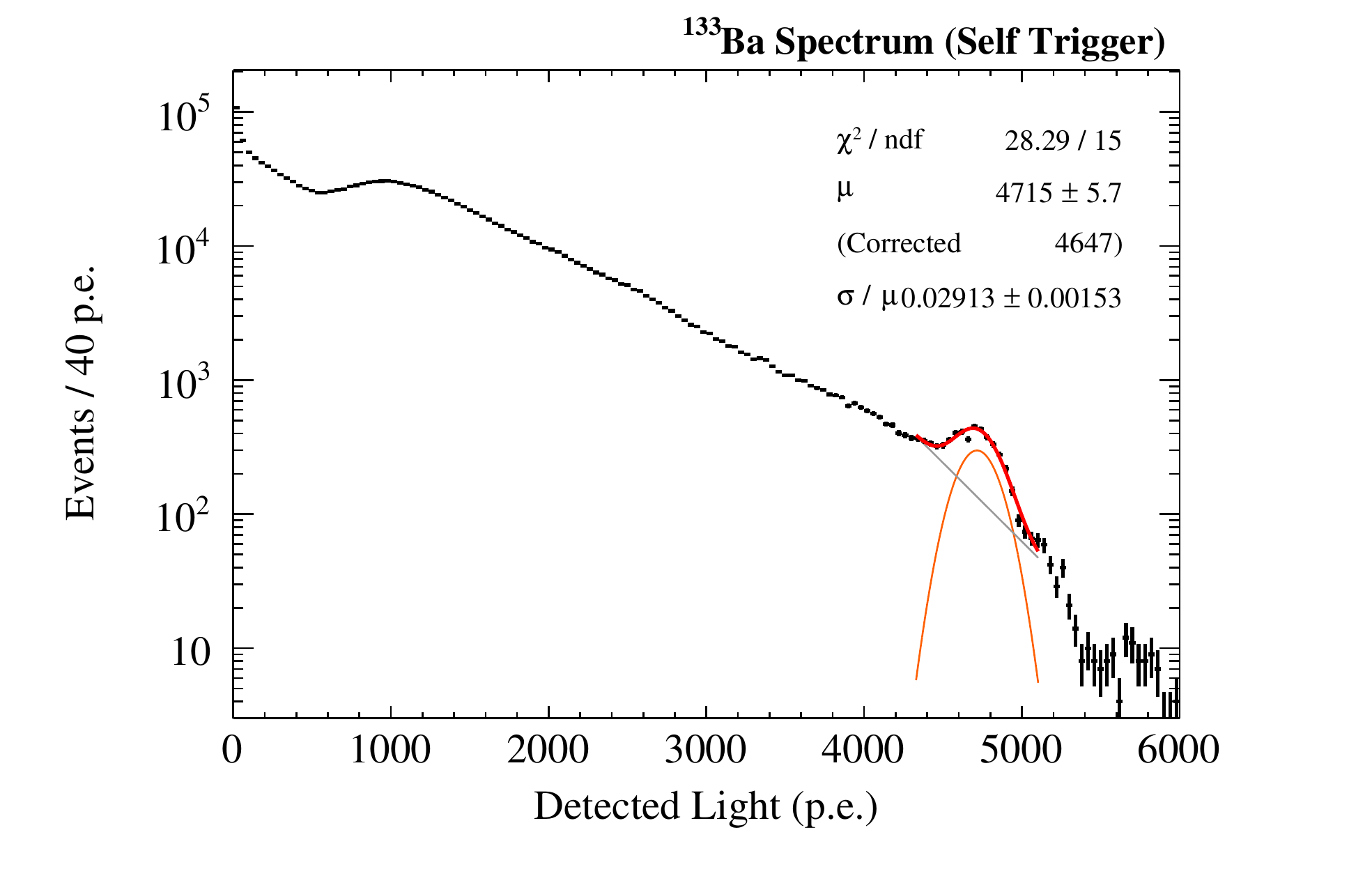}
	\caption{The observed light spectrum from the \riBa source.
	The red line represents the fit function.}
	\label{fig:133Ba_LY}
\end{figure}

The detector is exposed to a $356.0$-$\keV$ $\gamma$ ray using a \riBa radioactive source with approximately $1~\MBq$.
The spectrum obtained with a \riBa source is shown in Fig.~\ref{fig:133Ba_LY}.
The peak around $4700~\pe$ corresponds to the \gray line and fitted with a Gaussian.
An exponential function is added to the fit function to model the overall background components;
the main background sources are due to the degraded \gray tail and the \gray spectra of the other two lines of the \riBa source around the peak energy (those at $383.9$ and $302.9~\keV$) that have relatively high intensity.
The resulting fit function is overlaid in Fig.~\ref{fig:133Ba_LY}.

\subsection{Californium-252 source exploiting $\gamma$ rays through the $(n, n'\gamma)$ reaction with fluorine-19}
\label{subsec:Measurement_252Cf}
\begin{figure}[tb]
	\centering
	\includegraphics[width=\currentfigwid\columnwidth]{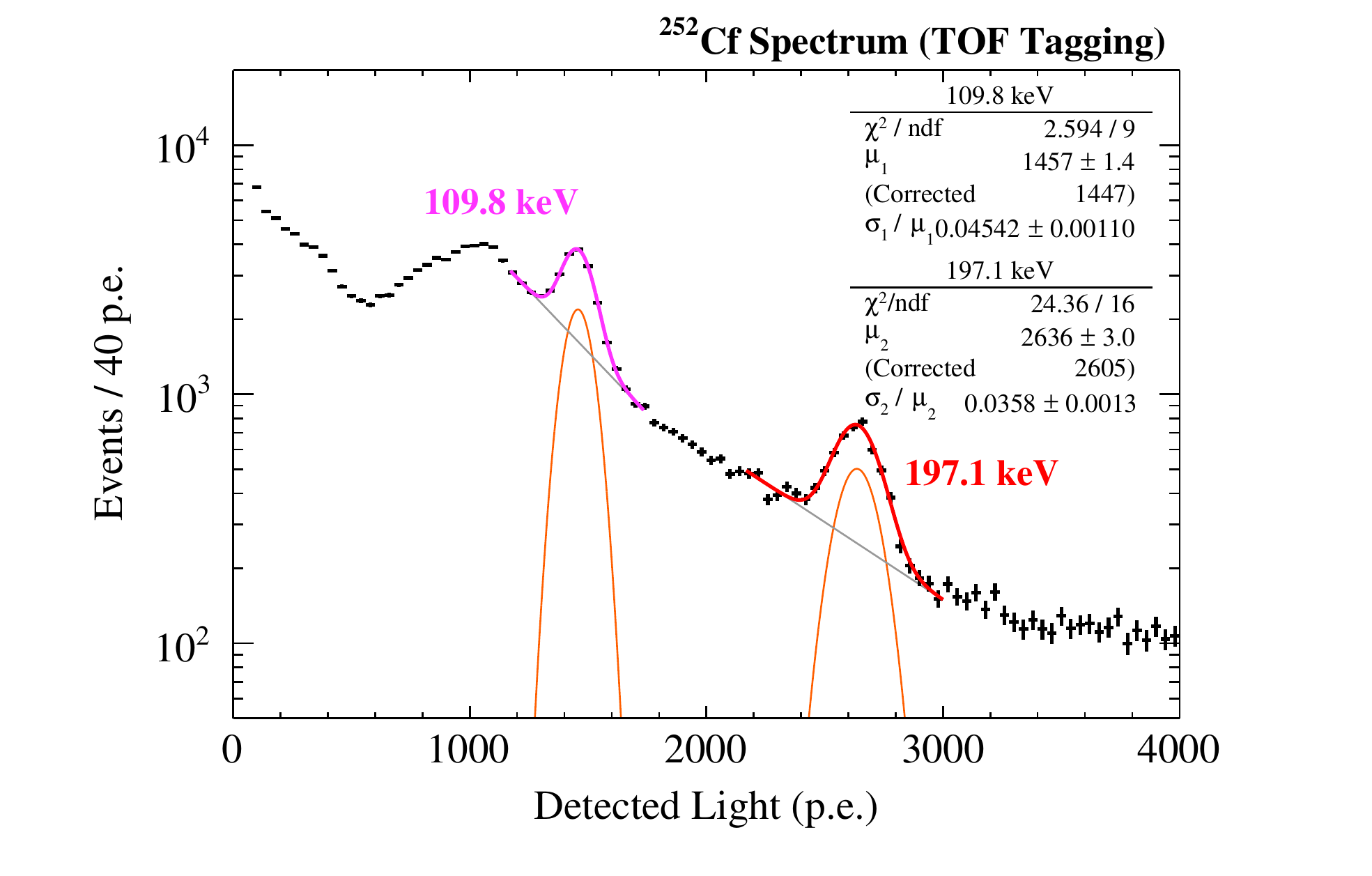}
	\caption{The observed light spectrum from the \riCf source after requiring the TOF to be consistent with fast neutrons.
	The magenta and red lines represent the fit functions for $109.8$- and $197.1$-$\keV$ peaks, respectively.}
	\label{fig:252Cf_LY}
\end{figure}

Measurements for the $109.8$- and $197.1$-$\keV$ quasimonoenergetic lines are performed using $\gamma$ rays emitted from the $(n, n'\gamma)$ reaction with \atomF \cite{rogers1974Inelastic}.
As an external fast neutron source, a \riCf source with a spontaneous fission rate of approximately $1\times10^5~{\rm fission/s}$ is used.
The distance between the center of the fiducial volume and the source is set to $90~\cm$.
The NaI(Tl) scintillator is placed beside the source to detect associated $\gamma$ rays from the spontaneous fission and to provide timing information.
Fast neutrons from \riCf generate $(n, n'\gamma)$ reaction with \atomF in the PTFE bulk, producing quasimonoenergetic $\gamma$ rays.
Although the intensities of each quasimonoenergetic line depend upon their incident neutron energy, $109.8$- and $197.1$-$\keV$ lines are major channels for the range of neutron energy from \riCf.
Time differences between the NaI(Tl) and fiducial signals (time of flight; TOF) are used to remove \gray events that come directly from the fission.
Figure~\ref{fig:252Cf_LY} shows the spectrum and fitting results for corresponding peaks.
Each peak is fit by a Gaussian plus exponential function.

\subsection{Americium-241 source}
\label{subsec:Measurement_241Am}
\begin{figure}[tb]
	\centering
	\includegraphics[width=\currentfigwid\columnwidth]{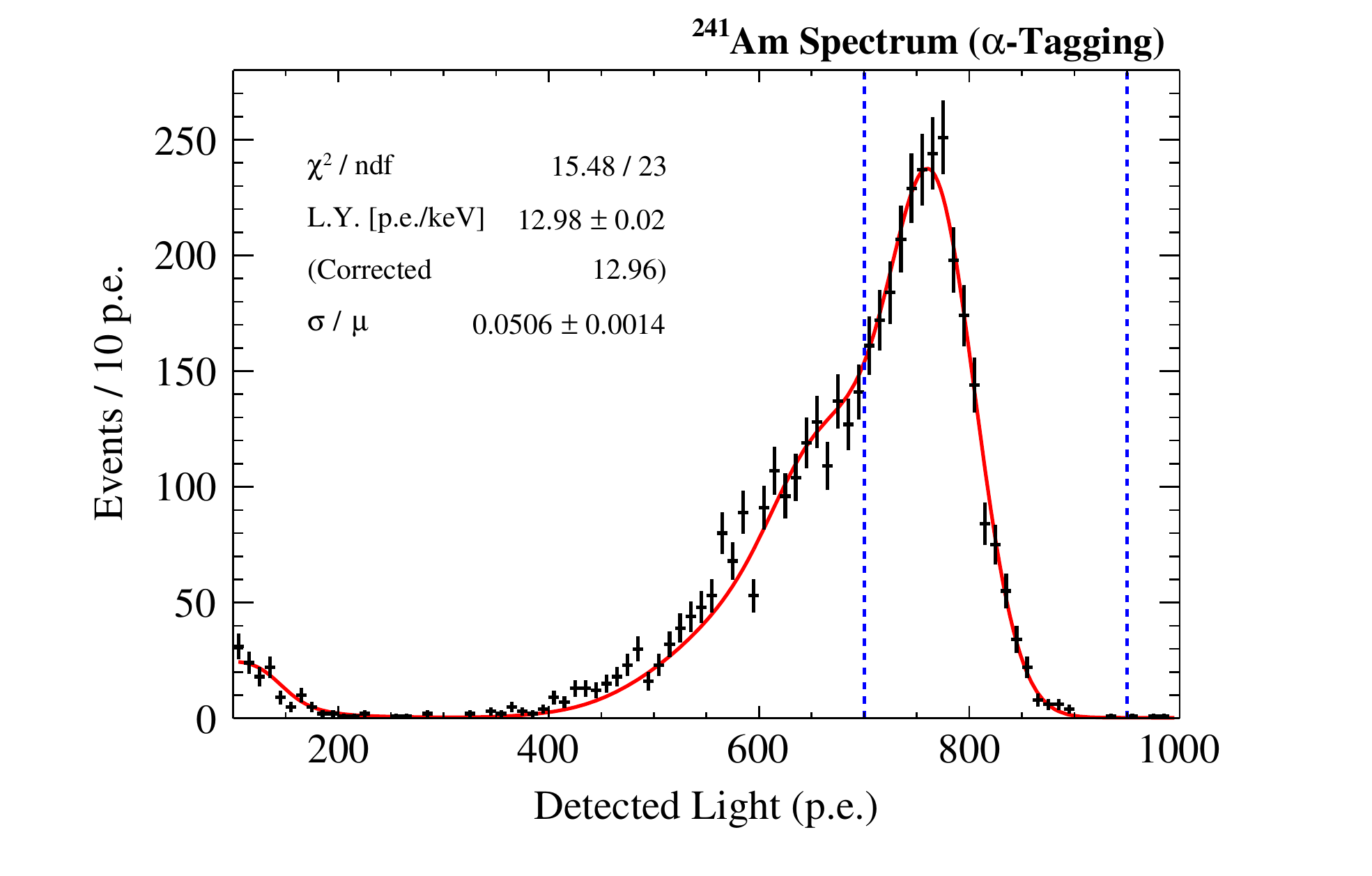}
	\caption{The observed light spectrum from the \riAm source by requiring \aray detection by the veto PMTs, along with the MC fit spectrum (red line).
	The blue dashed vertical lines represent the fitting range.}
	\label{fig:241Am_LY}
\end{figure}
To expose the detector to $59.5$-$\keV$ $\gamma$ rays, an \riAm source of approximately $40~\Bq$ is used.
The radioactive source is deposited on a $100$-$\um$-thick platinum foil installed at the outer surface of the PTFE bulk.
It decays into an excited level of $^{237}{\rm Np}$ via \aray transition, and subsequent deexcitation of the $^{237}{\rm Np}$ emits $\gamma$ rays with a major line of $59.5~\keV$.
The scintillation signal from the $\alpha$ ray from the primary disintegration is detected by the outer-bath PMTs, allowing the \gray interaction to be proved in the fiducial volume.
Figure~\ref{fig:241Am_LY} shows the observed light spectrum after requiring the detection of the \aray signals in the outer region.
Because of the relatively low energy of the $\gamma$ ray from \riAm and the passive components between the source and the fiducial volume, the spectrum does not exhibit a clear full-absorption peak.
The tail of the peak comes from $\gamma$ rays that reach the fiducial volume via single or multiple scattering from any materials in their path.

The detector response to a $59.5$-$\keV$ $\gamma$ ray is evaluated via MC simulation of the experimental setup based on the Geant4 toolkit \cite{agostinelli2003geant4, allison2006geant4}.
The MC simulation takes into account the detector geometry and composition inside the LAr bath, as well as the radioisotope mounting structure.
It proceeds by generating $\gamma$ rays from \riAm with a random momentum direction and calculating the energy deposition in the fiducial volume.
The observed spectrum is fitted by converting the energy deposition to the observed light yield with a constant scintillation yield, constant LCE, and Gaussian resolution.
The best fit spectrum is also shown in Fig.~\ref{fig:241Am_LY};
although the fit is performed only around the $59.5$-$\keV$ peak ($700$--$900~\pe$), reasonable agreement between the data and MC is found down to around $400~\pe$

\subsection{Argon-37 source}
\label{subsec:Measurement_37Ar}
\begin{figure}[tb]
	\centering
	\includegraphics[width=\currentfigwid\columnwidth]{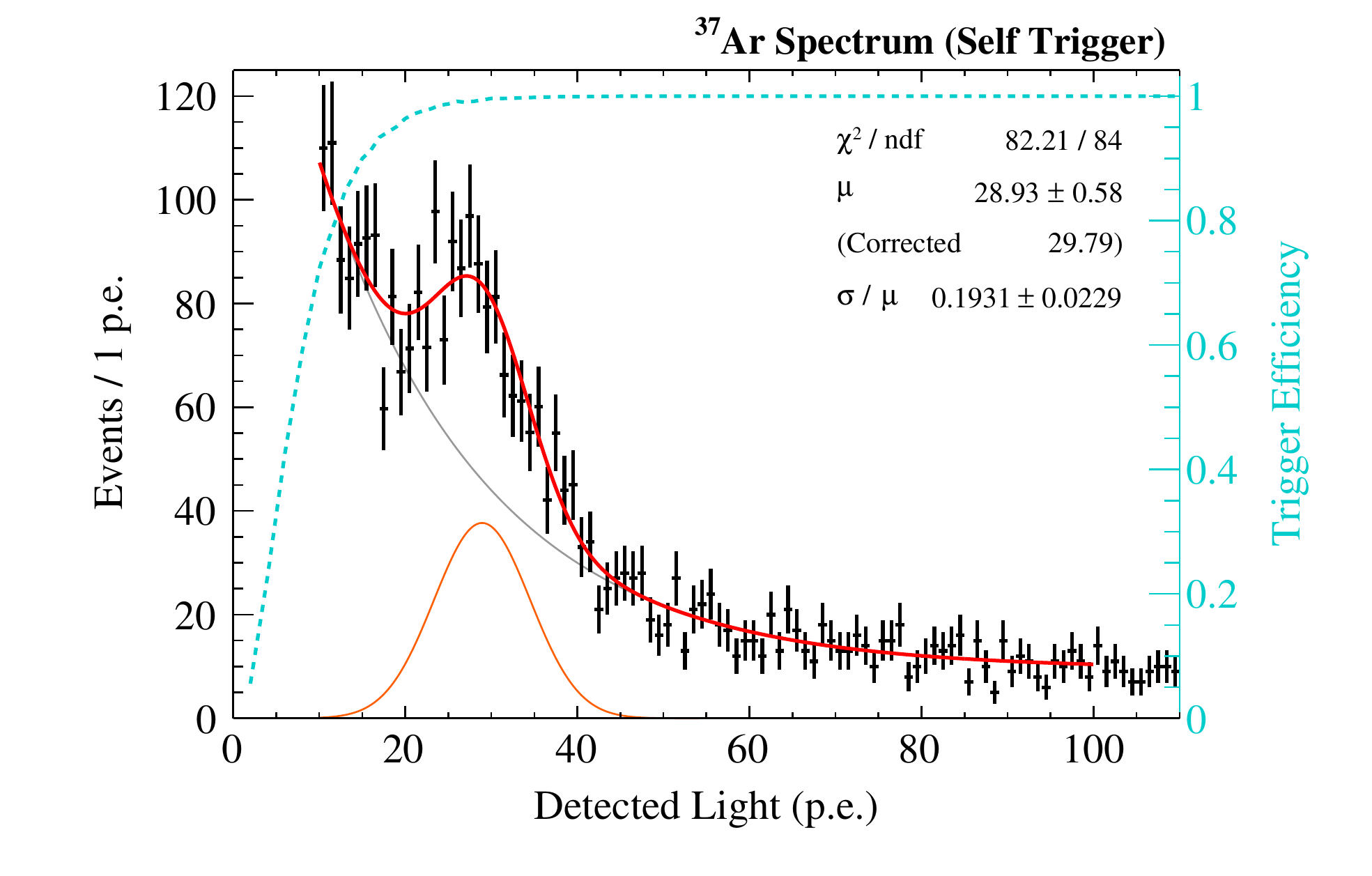}
	\caption{The \riArTS spectrum obtained by requiring anticoincidence with the outer bath PMTs for the no external source data.
	The cyan dashed line represents the estimated trigger efficiency, and the data are corrected based on this curve.}
	\label{fig:37Ar_LY}
\end{figure}

Measurement for ERs of a few $\keV$ is performed using \riArTS, which is the second most abundant radioactive isotope in atmospheric argon, comprising an abundance of $\approx$$1.3 \times 10^{-20}$ \cite{saldanha2019cosmogenic}.
It decays via electron capture to the ground state of $^{37}{\rm Cl}$ with a half-life of $35~{\rm days}$, producing x rays and Auger electrons with a total energy release of $2.82~\keV$ (for K-shell capture), $0.27~\keV$ (for L-shell capture), or $0.02~\keV$ (for M-shell capture) \cite{cleveland1998measurement, TabRad_v7}.
Since the production of \riArTS is mainly due to cosmogenic activation of atmospheric argon \cite{saldanha2019cosmogenic}, it is expected to reach equilibrium and the decay rate of \riArTS in the detector is expected to be constant from the argon filling time to the end of measurement.

The data used in this measurement come from approximately $27~{\rm hours}$ of detector operation without any external sources. 
Figure \ref{fig:37Ar_LY} shows the observed light spectrum for this measurement.
The spectrum consists of events that do not have associated scintillation signals in any of the four outer-bath PMTs.
The peak around $25~\pe$ is attributed to the energy release of $2.82~\keV$ from \riArTS.
No structures corresponding to the L- or M-shell capture could be seen, probably due to the large amount of random coincidence background and the lack of photostatistics.
The spectrum with \riArTS is fitted with the sum of the Gaussian, exponential, and constant terms that describe the signal and low energy background model.
The rate of \riArTS decays returned by the fit is approximately $25~{\rm mBq/kg}$, which is compatible with literature values \cite{saldanha2019cosmogenic, benetti2017Ar, agnes2018Low}.
The goodness of fit for the peak is $\chi^2/ndf=82.21/84$.

\section{Scintillation yield and energy resolution}
\label{sec:Result}

\begin{figure}[tb]
	\centering
	\includegraphics[width=\currentfigwid\columnwidth]{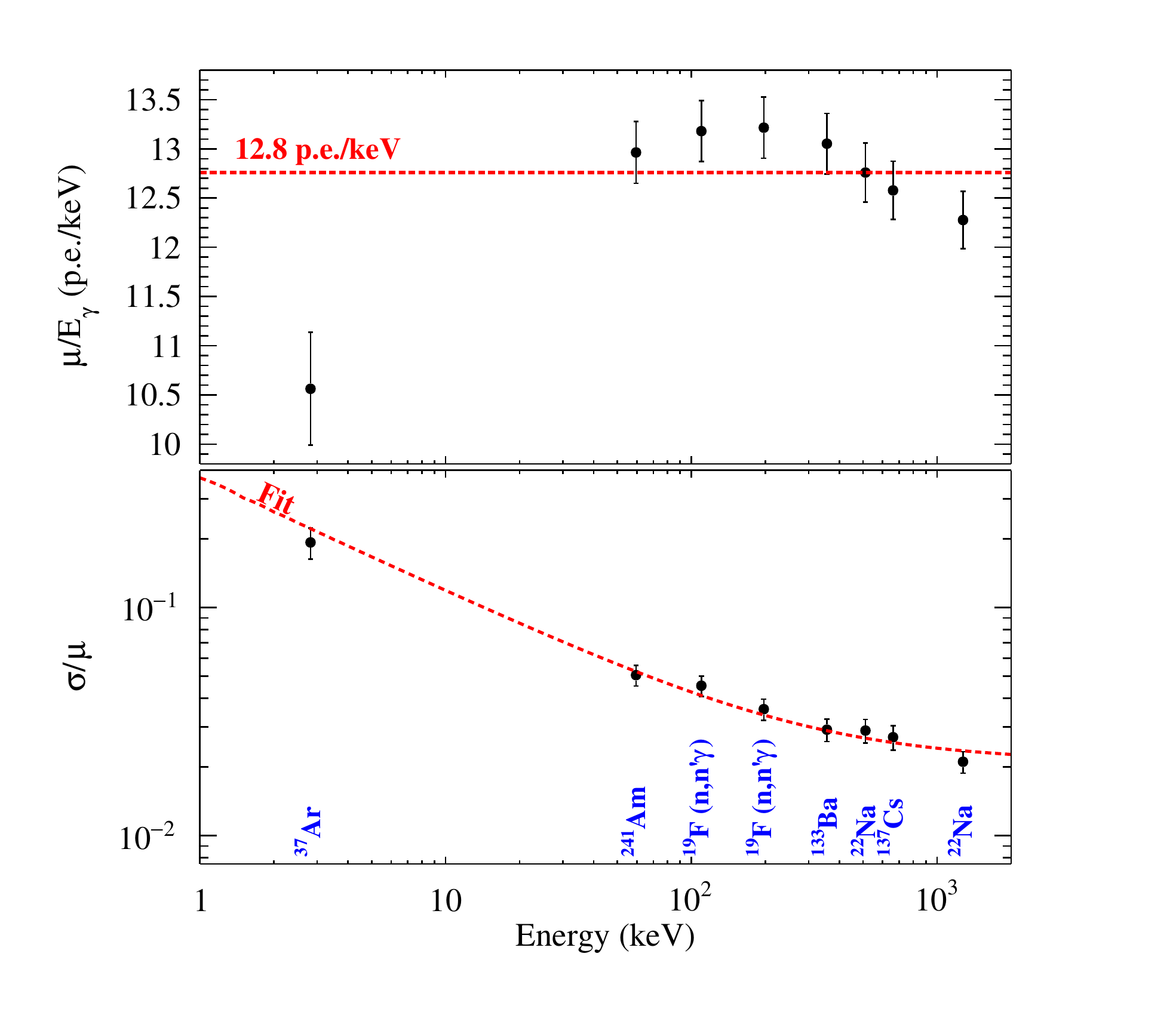}
	\caption{Top: observed light yields obtained by the fitting analysis for each calibration line divided by corresponding incident energy.
	The red dashed line represents the energy calibration using $511.0$-$\keV$ full-absorption peak.
	Bottom: energy resolution of the detector measured with full-absorption peaks.
	The red dashed line represents the fit function with stochastic and constant terms (see text).}	
	\label{fig:MeasSummary}
\end{figure}

\begin{table}[tb]
	\caption{Observed coefficients and estimated contributions of the stochastic (S) and constant (C) terms of the energy resolution.
	Although the origin of the constant term is not quantitatively estimated, almost all of which is believed to come from the geometrical effect.}
	\begin{tabular*}{\columnwidth}{@{\extracolsep{\fill}}lll} \hline \hline
		Type 								& Source					& Coefficient ($\alpha$) \\ \hline
		\multirow{8}{*}{\parbox[t]{0.2\columnwidth}{S \\ ($\frac{\sigma}{\mu} = \frac{\alpha}{\sqrt{E_\gamma}}$)}}	
											& Data					& $0.37 \pm 0.03$\\ \cline{2-3}
											& Photostatistics			& $\approx$$0.3$ \\	
											& Multiple scattering			& $<$$0.1$ \\
											& \multirow{2}{*}{\parbox[t]{0.3\columnwidth}{PMT gain and \newline afterpulse}}	& \multirow{2}{*}{$\lesssim$$0.2$} \\
											\\
											& \multirow{2}{*}{\parbox[t]{0.3\columnwidth}{Photoncounting \newline algorithm}}	& \multirow{2}{*}{$\approx$$0.0$} \\
											\\
											& TPB wavelength shift		& $0.0$--$0.1$ \\ \hline
		\multirow{2}{*}{C ($\frac{\sigma}{\mu} = \alpha$)}					& Data					& $0.021 \pm 0.002$ \\ \cline{2-3}
											& Geometrical effect			& ($\approx$$0.02$) \\ \hline \hline	
	\end{tabular*}
	\label{tab:ERes}
\end{table}

\begin{table*}[tb]
	\caption{Summary of the systematic uncertainty sources for the measurements of the light yields for each full-absorption peak and energy resolution.}
	\begin{tabular*}{0.8\hsize}{@{\extracolsep{\fill}}lllll} \hline \hline
		\multirow{2}{*}{Systematic}		& \multicolumn{2}{c}{Scintillation yields}		& \multicolumn{2}{c}{Energy resolution}	\\ \cline{2-3} \cline{4-5}
									& Dataset		& Fraction 			& Dataset		& Fraction		\\ \hline
		PMT afterpulse					& All			& $2.0\%$ 			& & \\		
		PMT gain nonlinearity			& All			& $<$$1.0\%$ 			& & \\		
		Time stability of the detector		& All			& $0.5\%$ 			& & \\
		Photon-counting algorithm			& All			& $1.0\%$ 			& & \\
		\multirow{2}{*}{Function modeling}	& \riAm		& $0.8\%$				& \multirow{2}{*}{All}	& \multirow{2}{*}{$10\%$} \\
									& Others		& $0.5\%$				& & \\ 
		\multirow{2}{*}{Trigger efficiency} 	& \riArTS		& $4.5\%$				& & \\
									& Others		& 0 					& & \\ \hline \hline
	\end{tabular*}
	\label{tab:LYUnc}
\end{table*}

The upper panel of Fig.~\ref{fig:MeasSummary} summarizes the mean values of the number of detected photoelectron divided by corresponding incident energies measured by the set of radioactive sources described in the previous section.
Nonlinear response on the scintillation yield is seen, which peaks around $200~\keV$.
This trend can be attributed to the energy dependence of the ionization electron-ion recombination probability.
The Thomas-Imel box (TIB) model \cite{thomas1987recombination} and Doke-Birks's law \cite{doke1988let} can presumably explain the data, as is the case for the liquid xenon (LXe) scintillation detector \cite{szydagis2011nest}.
For the higher energy range, the Doke-Birks's law is generally applied to deal with relativistic and longer-range tracks and to predict the decrease of the probability as the track energy increases (or $dE/dx$ decreases).
On the other hand, for the lower energy range, typically less than $\mathcal{O}(10~\keV)$, it is known that the TIB model is suitable for modeling the data because it is based on the low energy recoiled track whose range is comparable to or shorter than the mean ionization electron-ion thermalization distance.
The TIB model predicts the increase of the probability as the track energy increase (or number of ionization electron-ion pair increase).
Further study for quantitative evaluation and modeling of the LAr response will be discussed in Sec.~\ref{sec:Discuss}.

The energy resolution of the detector is also characterized based on the full-absorption peaks and is shown in the lower panel of Fig.~\ref{fig:MeasSummary}.
The set of points is fit to the function
\begin{equation}
	\frac{\sigma}{\mu}=\sqrt{\frac{\sigma_s^2}{E_{\gamma}} + \sigma_c^2},
\end{equation}
where $\sigma_s$ accounts for stochastic fluctuation, and $\sigma_c$ accounts for the variance of the mean value of monoenergy deposition.
The values are found to be $\sigma_s = 0.37 \pm 0.03$ and $\sigma_c = 0.021 \pm 0.002$, respectively.

Several sources are expected to degrade the energy resolution.
The contribution of each source is examined and listed in Table~\ref{tab:ERes}.
Convoluting the stochastic terms ($\sigma_s / \sqrt{E_\gamma }$) listed in Table~\ref{tab:ERes} explains approximately $90\%$ of the stochastic term observed in the data.
The rest of the term possibly comes from fluctuations in the ionization electron-ion recombination process; 
detecting the charge yield would be necessary to fully address it.
The constant term ($\sigma_c$) is believed to mainly consist of the geometrical effect.

The result is subjected to several systematic uncertainty sources which stem from both the detector response and the analysis procedure, as listed in Table~\ref{tab:LYUnc}.
The former is the linearity of the PMT gain and its afterpulse explored by the PMT response study using both LAr data and a property measurement of the PMT after the LAr detector operation and the time stability of the detector complex monitored by the regular calibrations throughout the data collection period.
The latter mainly comes from the photon-counting algorithm part and the related correction of the analysis.
We assign the size of the correction as the uncertainty.
Relatively small uncertainty is attributed to the fit of the full-absorption peak, which is estimated by refitting the peak with a simple Gaussian function.
The trigger efficiency is an additional uncertainty source for the \riArTS line analysis.
We refit the peak without the correction, and assign the corresponding uncertainty as the variation between these results.

The uncertainty of the energy resolution is considered as typically $10\%$ in total, mainly from the fitting modeling.

\section{TIB model interpretation on scintillation response}
\label{sec:Discuss}
\begin{figure}[tb]
	\centering
	\includegraphics[width=\currentfigwid\columnwidth]{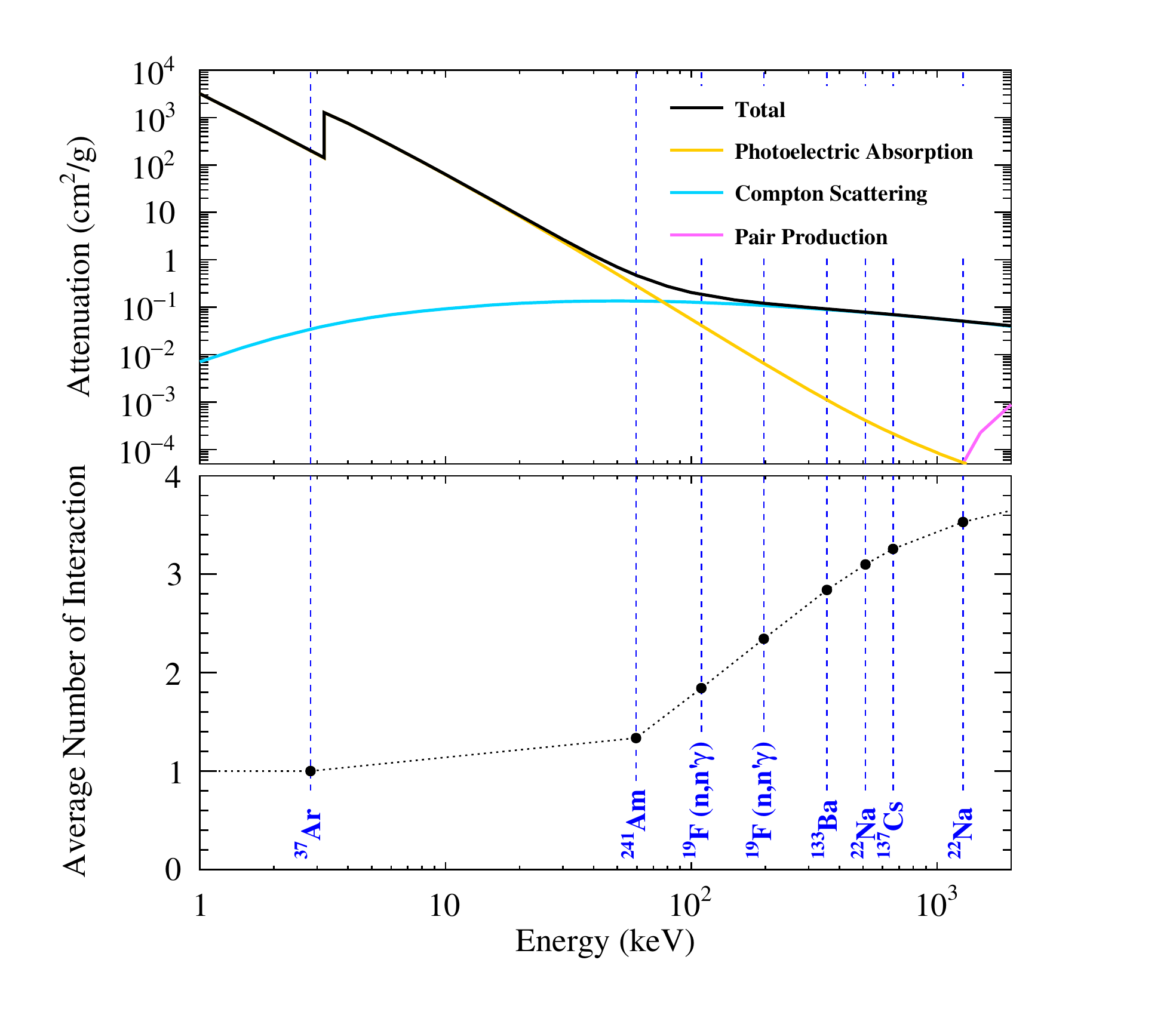}
	\caption{Top: the \gray cross sections for argon provided by XCOM \cite{xcom}.
	Bottom: average number of interaction points for the full-absorption peaks calculated by the Geant4 MC simulation.}
	\label{fig:G4truth}
\end{figure}
\begin{figure}[tb]
	\centering
	\includegraphics[width=\currentfigwid\columnwidth]{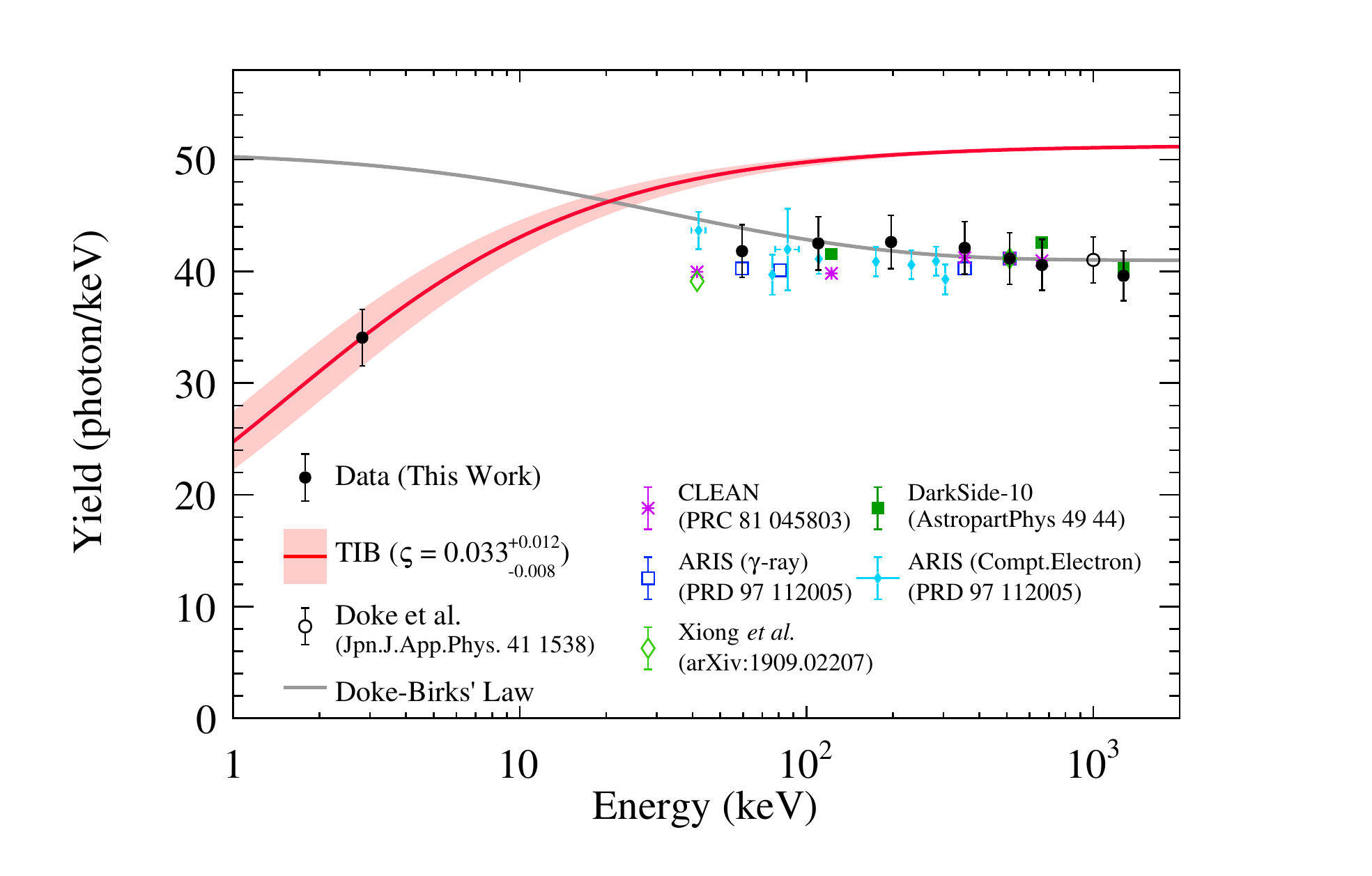}
	\caption{Measured scintillation yield as a function of the incident energy $E_\gamma$ (black solid circle).
	The absolute yield is determined by referring the measurement by Doke {\it et al.} (black open circle) \cite{Doke2002Absolute}.
	The TIB model function with a parameter found by the $2.82$-$\keV$ point is shown with its uncertainty (red band).
	The results from other experiments, CLEAN (violet star point) \cite{lippincott2010calibration}, DarkSide-10 (green filled square) \cite{Alexander2013light}, ARIS (blue open square and blue filled rhombus) \cite{agnes2018measurement}, and Xiong {\it et al.} (green open rhombus) \cite{Xiong2019calibration} are also shown where each yield is normalized at $511~\keV$ referring the Doke-Birks's law (gray solid line) \cite{Doke2002Absolute}.}
	\label{fig:NphSummary}
\end{figure}

The absolute scintillation yield, the number of photons generated by an incident particle $n_{ph}$ per unit energy deposition ${\rm photon}/\keV$ is a more essential quantity for the LAr scintillation detector than the observed light signal per incident energy $\pe/\keV$. 
The yield for a recoiled electron is measured by Doke {\it et al.} as $41 \pm 2~{\rm photon}/\keV$ using a $1$-$\MeV$ \bray source \cite{Doke2002Absolute}.

On the other hand, the $511$-$\keV$ full-absorption point is the most suitable energy for the comparison between the previous measurements since several works \cite{lippincott2010calibration, Alexander2013light, agnes2018measurement, Xiong2019calibration} have commonly presented the observed light yield at this point.
As the scintillation yield of a \gray full-absorption event is affected by the energy dependence of that for recoiled electrons because of multiple scattering, 
we perform a Geant4 MC simulation to evaluate it.
Figure~\ref{fig:G4truth} shows the average number of the interaction points.
It indicates that $511$-$\keV$ full-absorption events contain about three interaction points on average;
however, a discrepancy between the yield for the $511$-$\keV$ $\gamma$ ray and that for the $\beta$ ray is found to be less than $2\%$ when assuming Doke-Birks's law (gray line in Fig.~\ref{fig:NphSummary}) \cite{Doke2002Absolute}.
Therefore, we determine the absolute scintillation yield by using the $511$-$\keV$ point and referring the Doke's measurement.

Figure~\ref{fig:NphSummary} shows the scintillation yield obtained in this analysis.
As mentioned in the previous section, the energy dependence of the yield is attributed to the ionization electron-ion recombination probability.
For a lower energy event, the TIB model presumably predicts the response
\begin{eqnarray}
	\label{eq:Nph_TIB}
	n_{ph} &=& \frac{E_{\rm er}}{W} (N_{ex} + r N_i) = \frac{E_{\rm er}}{W} \frac{1+r}{1+\alpha}, \\ \nonumber
	r &=& 1 - \frac{1}{N_i\varsigma}\ln (1+N_i \varsigma),
\end{eqnarray}
where $E_{\rm er}$ is the recoiled electron energy, $W = 19.5~\eV$ is the effective work function \cite{doke1988let}, $N_{ex}$ and $N_i$ are the numbers of produced excitons or electron-ion pairs, respectively, $\alpha = 0.21$ is the initial ratio of the average of $N_{ex}$ to $N_i$ \cite{miyajima1974average}, and $\varsigma$ is a constant parameter of the model.
Considering the facts that the number of the interaction point of the \riArTS events can be approximated to be one due to its low energy deposition and decay mode mainly consisting of Auger electrons \cite{cleveland1998measurement}, and that the TIB model is fully applied for liquid xenon at corresponding energy where the electron track length is smaller than the thermalization distance of the ionization electron \cite{szydagis2011nest, Mozumder1995Free}, we determine the parameter $\varsigma$ from the \riArTS data. 
It is calculated as $\varsigma=0.033 ^{+0.012} _{-0.008}$ and represented with the red band in Fig.~\ref{fig:NphSummary}.
Further studies, such as additional measurements around $10~\keV$ and discussion on the stitching between the TIB model and Doke-Birks's law, should be performed in future work.
This result also would be practically essential input for tuning the response model implemented, for instance, in the NEST package \cite{nestweb}.

\section{Conclusion}
\label{sec:Conclusion}
The energy dependence of the scintillation yield for electronic recoils ranging from $2.82$ to $1274.6~\keV$ is measured using a single-phase detector with high LCE exposed to a variety of calibration sources.
The scintillation detector with the TPB wavelength shifter is immersed in purified LAr and yields $12.8 \pm 0.3~\pe/\keV$ ($11.2 \pm 0.3~\pe/\keV$) for a $511.0$-$\keV$ \gray full-absorption event based on the PMT calibration assuming a PMT single photoelectron response model with an additional exponential term (with only a Gaussian term), and its energy resolution is $3\%$ for the \gray line.
The scintillation response is investigated by the full-absorption peaks of external \gray sources, as well as an \riArTS source with a $2.82$-$\keV$ line.
These measurements demonstrate that the scintillation yield decreases in the low energy region.
We interpret it by analogy with the LXe scintillation detector response, where the ionization electron-ion recombination probability is attributed to the energy dependence of the yield.
By referring the previous measurement of the scintillation yield at $1~\MeV$, the TIB model parameter $\varsigma$ is calculated by the $2.82$-$\keV$ point as $\varsigma = 0.033 ^{+0.012}_{-0.008}$.

This work is primarily intended for use in the direct WIMP dark matter search.
In this field, low energy electronic background is one of the most severe sources disturbing the lower energy threshold, hence, reducing WIMP sensitivity.
The result presented here makes use of the precise estimation of background contamination in the low energy region and suppression of the systematic uncertainty.
The measurement of the scintillation response under nonzero electric field, which is the matter for a double-phase detector (e.g., \cite{agnes2018darkside, Aalseth2017DarkSide}), is left for future work.
In addition, the measurement of the energy resolution for the $\keV$ to $\MeV$ range in this work provides useful information for applying the LAr detector to other fields, such as astrophysical MeV gamma-ray observation \cite{aramaki2020dual}.
The results presented here would help with the design, operation, and analysis of a wide variety of astrophysical and particle physics experiments in the near future to enhance their physical reach.

\section*{Acknowledgments}
\label{sec:Acknow}

This work is a part of the outcome of research performed under the Waseda University Research Institute for Science and Engineering (Project No. 2016A-507) supported by Japan Society for the Promotion of Science Grant-in-Aid for Scientific Research on Innovative Areas (Grants No. 15H01038 and No. 17H05204), Grant-in-Aid for Scientific Research(B) (Grant No. 18H01234), and Grant-in-Aid for Japan Society for the Promotion of Science Research Fellow (Grant No. 18J13018).
The authors would like to thank the Material Characterization Central Laboratory at Waseda University for granting us access to their stylus profiler.
The authors acknowledge the support of the Institute for Advanced Theoretical and Experimental Physics, Waseda University.

\nocite{*}

\bibliography{Kimura}

\providecommand{\noopsort}[1]{}\providecommand{\singleletter}[1]{#1}%
\begin{thebibliography}{37}
\expandafter\ifx\csname natexlab\endcsname\relax\def\natexlab#1{#1}\fi
\expandafter\ifx\csname bibnamefont\endcsname\relax
  \def\bibnamefont#1{#1}\fi
\expandafter\ifx\csname bibfnamefont\endcsname\relax
  \def\bibfnamefont#1{#1}\fi
\expandafter\ifx\csname citenamefont\endcsname\relax
  \def\citenamefont#1{#1}\fi
\expandafter\ifx\csname url\endcsname\relax
  \def\url#1{\texttt{#1}}\fi
\expandafter\ifx\csname urlprefix\endcsname\relax\def\urlprefix{URL }\fi
\providecommand{\bibinfo}[2]{#2}
\providecommand{\eprint}[2][]{\url{#2}}

\bibitem[{\citenamefont{Agnes et~al.}(2018{\natexlab{a}})\citenamefont{Agnes,
  Albuquerque, Alexander, Alton, Araujo, Ave, Back, Baldin, Batignani, Biery
  et~al.}}]{agnes2018darkside}
\bibinfo{author}{\bibfnamefont{P.}~\bibnamefont{Agnes}},
  \bibinfo{author}{\bibfnamefont{I.~F.~M.} \bibnamefont{Albuquerque}},
  \bibinfo{author}{\bibfnamefont{T.}~\bibnamefont{Alexander}},
  \bibinfo{author}{\bibfnamefont{A.~K.} \bibnamefont{Alton}},
  \bibinfo{author}{\bibfnamefont{G.~R.} \bibnamefont{Araujo}},
  \bibinfo{author}{\bibfnamefont{M.}~\bibnamefont{Ave}},
  \bibinfo{author}{\bibfnamefont{H.~O.} \bibnamefont{Back}},
  \bibinfo{author}{\bibfnamefont{B.}~\bibnamefont{Baldin}},
  \bibinfo{author}{\bibfnamefont{G.}~\bibnamefont{Batignani}},
  \bibinfo{author}{\bibfnamefont{K.}~\bibnamefont{Biery}}, \bibnamefont{et~al.}
  (\bibinfo{collaboration}{DarkSide Collaboration}),
  \bibinfo{title}{{D}ark{S}ide-50 532-day dark matter search with low-radioactivity argon},
  \bibinfo{journal}{Phys.
  Rev. D} \textbf{\bibinfo{volume}{98}}, \bibinfo{pages}{102006}
  (\bibinfo{year}{2018}{\natexlab{a}}).

\bibitem[{\citenamefont{Ajaj et~al.}(2019)\citenamefont{Ajaj, Amaudruz, Araujo,
  Baldwin, Batygov, Beltran, Bina, Bonatt, Boulay, Broerman
  et~al.}}]{ajaj2019search}
\bibinfo{author}{\bibfnamefont{R.}~\bibnamefont{Ajaj}},
  \bibinfo{author}{\bibfnamefont{P.-A.} \bibnamefont{Amaudruz}},
  \bibinfo{author}{\bibfnamefont{G.~R.} \bibnamefont{Araujo}},
  \bibinfo{author}{\bibfnamefont{M.}~\bibnamefont{Baldwin}},
  \bibinfo{author}{\bibfnamefont{M.}~\bibnamefont{Batygov}},
  \bibinfo{author}{\bibfnamefont{B.}~\bibnamefont{Beltran}},
  \bibinfo{author}{\bibfnamefont{C.~E.} \bibnamefont{Bina}},
  \bibinfo{author}{\bibfnamefont{J.}~\bibnamefont{Bonatt}},
  \bibinfo{author}{\bibfnamefont{M.~G.} \bibnamefont{Boulay}},
  \bibinfo{author}{\bibfnamefont{B.}~\bibnamefont{Broerman}},
  \bibnamefont{et~al.} (\bibinfo{collaboration}{DEAP Collaboration}),
  \bibinfo{title}{Search for dark matter with a 231-day exposure of liquid argon using {DEAP}-3600 at {SNOLAB}},
  \bibinfo{journal}{Phys. Rev. D} \textbf{\bibinfo{volume}{100}},
  \bibinfo{pages}{022004} (\bibinfo{year}{2019}).
  
\bibitem[{\citenamefont{Akerib et~al.}(2017)\citenamefont{Akerib, Alsum,
  Aquino, Ara\'ujo, Bai, Bailey, Balajthy, Beltrame, Bernard, Bernstein
  et~al.}}]{akerib2017first}
\bibinfo{author}{\bibfnamefont{D.~S.} \bibnamefont{Akerib}},
  \bibinfo{author}{\bibfnamefont{S.}~\bibnamefont{Alsum}},
  \bibinfo{author}{\bibfnamefont{C.}~\bibnamefont{Aquino}},
  \bibinfo{author}{\bibfnamefont{H.~M.} \bibnamefont{Ara\'ujo}},
  \bibinfo{author}{\bibfnamefont{X.}~\bibnamefont{Bai}},
  \bibinfo{author}{\bibfnamefont{A.~J.} \bibnamefont{Bailey}},
  \bibinfo{author}{\bibfnamefont{J.}~\bibnamefont{Balajthy}},
  \bibinfo{author}{\bibfnamefont{P.}~\bibnamefont{Beltrame}},
  \bibinfo{author}{\bibfnamefont{E.~P.} \bibnamefont{Bernard}},
  \bibinfo{author}{\bibfnamefont{A.}~\bibnamefont{Bernstein}},
  \bibnamefont{et~al.} (\bibinfo{collaboration}{LUX Collaboration}),
  \bibinfo{title}{First Searches for Axions and Axionlike Particles with the LUX Experiment},
  \bibinfo{journal}{Phys. Rev. Lett.} \textbf{\bibinfo{volume}{118}},
  \bibinfo{pages}{261301} (\bibinfo{year}{2017}).
  
\bibitem[{\citenamefont{Aprile et~al.}(2019)\citenamefont{Aprile, Aalbers,
  Agostini, Alfonsi, Althueser, Amaro, Antochi, Angelino, Arneodo, Barge
  et~al.}}]{aprile2019light}
\bibinfo{author}{\bibfnamefont{E.}~\bibnamefont{Aprile}},
  \bibinfo{author}{\bibfnamefont{J.}~\bibnamefont{Aalbers}},
  \bibinfo{author}{\bibfnamefont{F.}~\bibnamefont{Agostini}},
  \bibinfo{author}{\bibfnamefont{M.}~\bibnamefont{Alfonsi}},
  \bibinfo{author}{\bibfnamefont{L.}~\bibnamefont{Althueser}},
  \bibinfo{author}{\bibfnamefont{F.~D.} \bibnamefont{Amaro}},
  \bibinfo{author}{\bibfnamefont{V.~C.} \bibnamefont{Antochi}},
  \bibinfo{author}{\bibfnamefont{E.}~\bibnamefont{Angelino}},
  \bibinfo{author}{\bibfnamefont{F.}~\bibnamefont{Arneodo}},
  \bibinfo{author}{\bibfnamefont{D.}~\bibnamefont{Barge}}, \bibnamefont{et~al.}
  (\bibinfo{collaboration}{XENON Collaboration}),
  \bibinfo{title}{Light Dark Matter Search with Ionization Signals in XENON1T},
  \bibinfo{journal}{Phys. Rev.
  Lett.} \textbf{\bibinfo{volume}{123}}, \bibinfo{pages}{251801}
  (\bibinfo{year}{2019}).

\bibitem[{\citenamefont{Aprile et~al.}(2020)}]{Aprile2020Observation}
\bibinfo{author}{\bibfnamefont{E.}~\bibnamefont{Aprile}} \bibnamefont{et~al.}
(\bibinfo{collaboration}{XENON Collaboration}),
\bibinfo{title}{Observation of Excess Electronic Recoil Events in XENON1T},
(\bibinfo{year}{2020}),
  \eprint{2006.09721}.

\bibitem[{\citenamefont{Szydagis et~al.}(2011)\citenamefont{Szydagis, Barry,
  Kazkaz, Mock, Stolp, Sweany, Tripathi, Uvarov, Walsh, and
  Woods}}]{szydagis2011nest}
\bibinfo{author}{\bibfnamefont{M.}~\bibnamefont{Szydagis}},
  \bibinfo{author}{\bibfnamefont{N.}~\bibnamefont{Barry}},
  \bibinfo{author}{\bibfnamefont{K.}~\bibnamefont{Kazkaz}},
  \bibinfo{author}{\bibfnamefont{J.}~\bibnamefont{Mock}},
  \bibinfo{author}{\bibfnamefont{D.}~\bibnamefont{Stolp}},
  \bibinfo{author}{\bibfnamefont{M.}~\bibnamefont{Sweany}},
  \bibinfo{author}{\bibfnamefont{M.}~\bibnamefont{Tripathi}},
  \bibinfo{author}{\bibfnamefont{S.}~\bibnamefont{Uvarov}},
  \bibinfo{author}{\bibfnamefont{N.}~\bibnamefont{Walsh}}, \bibnamefont{and}
  \bibinfo{author}{\bibfnamefont{M.}~\bibnamefont{Woods}},
  \bibinfo{title}{{NEST}: a comprehensive model for scintillation yield in liquid xenon},
  \bibinfo{journal}{J.
  Instrum.} \textbf{\bibinfo{volume}{6}}, \bibinfo{pages}{P10002}
  (\bibinfo{year}{2011}).

\bibitem[{\citenamefont{Lenardo et~al.}(2015)\citenamefont{Lenardo, Kazkaz,
  Manalaysay, Mock, Szydagis, and Tripathi}}]{lenardo2015global}
\bibinfo{author}{\bibfnamefont{B.}~\bibnamefont{Lenardo}},
  \bibinfo{author}{\bibfnamefont{K.}~\bibnamefont{Kazkaz}},
  \bibinfo{author}{\bibfnamefont{A.}~\bibnamefont{Manalaysay}},
  \bibinfo{author}{\bibfnamefont{J.}~\bibnamefont{Mock}},
  \bibinfo{author}{\bibfnamefont{M.}~\bibnamefont{Szydagis}}, \bibnamefont{and}
  \bibinfo{author}{\bibfnamefont{M.}~\bibnamefont{Tripathi}},
  \bibinfo{title}{A Global Analysis of Light and Charge Yields in Liquid Xenon},
  \bibinfo{journal}{IEEE Trans. Nucl. Sci.} \textbf{\bibinfo{volume}{62}},
  \bibinfo{pages}{3387} (\bibinfo{year}{2015}), ISSN \bibinfo{issn}{0018-9499}.

\bibitem[{\citenamefont{Szydagis et~al.}(2018)\citenamefont{Szydagis, Balajthy,
  Brodsky, Cutter, Huang, Kozlova, Lenardo, Manalaysay, McKinsey, Mooney
  et~al.}}]{szydagis2018noble}
\bibinfo{author}{\bibfnamefont{M.}~\bibnamefont{Szydagis}},
  \bibinfo{author}{\bibfnamefont{J.}~\bibnamefont{Balajthy}},
  \bibinfo{author}{\bibfnamefont{J.}~\bibnamefont{Brodsky}},
  \bibinfo{author}{\bibfnamefont{J.}~\bibnamefont{Cutter}},
  \bibinfo{author}{\bibfnamefont{J.}~\bibnamefont{Huang}},
  \bibinfo{author}{\bibfnamefont{E.}~\bibnamefont{Kozlova}},
  \bibinfo{author}{\bibfnamefont{B.}~\bibnamefont{Lenardo}},
  \bibinfo{author}{\bibfnamefont{A.}~\bibnamefont{Manalaysay}},
  \bibinfo{author}{\bibfnamefont{D.}~\bibnamefont{McKinsey}},
  \bibinfo{author}{\bibfnamefont{M.}~\bibnamefont{Mooney}},
  \bibnamefont{et~al.},
  \bibinfo{title}{Noble Element Simulation Technique v2.0},
  \bibinfo{doi}{10.5281/zenodo.1314669},
  (\bibinfo{year}{2018}).

\bibitem[{\citenamefont{Hitachi et~al.}(1983)\citenamefont{Hitachi, Takahashi,
  Funayama, Masuda, Kikuchi, and Doke}}]{hitachi1983effect}
\bibinfo{author}{\bibfnamefont{A.}~\bibnamefont{Hitachi}},
  \bibinfo{author}{\bibfnamefont{T.}~\bibnamefont{Takahashi}},
  \bibinfo{author}{\bibfnamefont{N.}~\bibnamefont{Funayama}},
  \bibinfo{author}{\bibfnamefont{K.}~\bibnamefont{Masuda}},
  \bibinfo{author}{\bibfnamefont{J.}~\bibnamefont{Kikuchi}}, \bibnamefont{and}
  \bibinfo{author}{\bibfnamefont{T.}~\bibnamefont{Doke}},
  \bibinfo{title}{Effect of ionization density on the time dependence of luminescence from liquid argon and xenon},
  \bibinfo{journal}{Phys. Rev. B} \textbf{\bibinfo{volume}{27}},
  \bibinfo{pages}{5279} (\bibinfo{year}{1983}).

\bibitem[{\citenamefont{Heindl et~al.}(2010)\citenamefont{Heindl, Dandl,
  Hofmann, Krücken, Oberauer, Potzel, Wieser, and Ulrich}}]{heindl2010the}
\bibinfo{author}{\bibfnamefont{T.}~\bibnamefont{Heindl}},
  \bibinfo{author}{\bibfnamefont{T.}~\bibnamefont{Dandl}},
  \bibinfo{author}{\bibfnamefont{M.}~\bibnamefont{Hofmann}},
  \bibinfo{author}{\bibfnamefont{R.}~\bibnamefont{Krücken}},
  \bibinfo{author}{\bibfnamefont{L.}~\bibnamefont{Oberauer}},
  \bibinfo{author}{\bibfnamefont{W.}~\bibnamefont{Potzel}},
  \bibinfo{author}{\bibfnamefont{J.}~\bibnamefont{Wieser}}, \bibnamefont{and}
  \bibinfo{author}{\bibfnamefont{A.}~\bibnamefont{Ulrich}},
  \bibinfo{title}{The scintillation of liquid argon},
  \bibinfo{journal}{Europhys. Lett.} \textbf{\bibinfo{volume}{91}},
  \bibinfo{pages}{62002} (\bibinfo{year}{2010}).

\bibitem[{\citenamefont{Burton and Powell}(1973)}]{burton1973fluorescence}
\bibinfo{author}{\bibfnamefont{W.~M.} \bibnamefont{Burton}} \bibnamefont{and}
\bibinfo{author}{\bibfnamefont{B.~A.} \bibnamefont{Powell}},
\bibinfo{title}{Fluorescence of Tetraphenyl-Butadiene in the Vacuum Ultraviolet},
  \bibinfo{journal}{Appl. Opt.} \textbf{\bibinfo{volume}{12}},
  \bibinfo{pages}{87} (\bibinfo{year}{1973}).

\bibitem[{\citenamefont{Porter and Topp}(1970)}]{porter1970nanosecond}
\bibinfo{author}{\bibfnamefont{G.}~\bibnamefont{Porter}} \bibnamefont{and}
\bibinfo{author}{\bibfnamefont{M.~R.} \bibnamefont{Topp}},
\bibinfo{title}{Nanosecond flash photolysis},
  \bibinfo{journal}{Proc. R. Soc. A} \textbf{\bibinfo{volume}{315}}, \bibinfo{pages}{163}
  (\bibinfo{year}{1970}).

\bibitem[{\citenamefont{Acciarri
  et~al.}(2010{\natexlab{a}})\citenamefont{Acciarri, Antonello, Baibussinov,
  Baldo-Ceolin, Benetti, Calaprice, Calligarich, Cambiaghi, Canci, Carbonara
  et~al.}}]{acciarri2010oxygen}
\bibinfo{author}{\bibfnamefont{R.}~\bibnamefont{Acciarri}},
  \bibinfo{author}{\bibfnamefont{M.}~\bibnamefont{Antonello}},
  \bibinfo{author}{\bibfnamefont{B.}~\bibnamefont{Baibussinov}},
  \bibinfo{author}{\bibfnamefont{M.}~\bibnamefont{Baldo-Ceolin}},
  \bibinfo{author}{\bibfnamefont{P.}~\bibnamefont{Benetti}},
  \bibinfo{author}{\bibfnamefont{F.}~\bibnamefont{Calaprice}},
  \bibinfo{author}{\bibfnamefont{E.}~\bibnamefont{Calligarich}},
  \bibinfo{author}{\bibfnamefont{M.}~\bibnamefont{Cambiaghi}},
  \bibinfo{author}{\bibfnamefont{N.}~\bibnamefont{Canci}},
  \bibinfo{author}{\bibfnamefont{F.}~\bibnamefont{Carbonara}},
  \bibnamefont{et~al.},
  \bibinfo{title}{Oxygen contamination in liquid Argon: combined effects on ionization electron charge and scintillation light},
  \bibinfo{journal}{J. Instrum.}
  \textbf{\bibinfo{volume}{5}}, \bibinfo{pages}{P05003}
  (\bibinfo{year}{2010}{\natexlab{a}}).

\bibitem[{\citenamefont{Acciarri
  et~al.}(2010{\natexlab{b}})\citenamefont{Acciarri, Antonello, Baibussinov,
  Baldo-Ceolin, Benetti, Calaprice, Calligarich, Cambiaghi, Canci, Carbonara
  et~al.}}]{acciarri2010effects}
\bibinfo{author}{\bibfnamefont{R.}~\bibnamefont{Acciarri}},
  \bibinfo{author}{\bibfnamefont{M.}~\bibnamefont{Antonello}},
  \bibinfo{author}{\bibfnamefont{B.}~\bibnamefont{Baibussinov}},
  \bibinfo{author}{\bibfnamefont{M.}~\bibnamefont{Baldo-Ceolin}},
  \bibinfo{author}{\bibfnamefont{P.}~\bibnamefont{Benetti}},
  \bibinfo{author}{\bibfnamefont{F.}~\bibnamefont{Calaprice}},
  \bibinfo{author}{\bibfnamefont{E.}~\bibnamefont{Calligarich}},
  \bibinfo{author}{\bibfnamefont{M.}~\bibnamefont{Cambiaghi}},
  \bibinfo{author}{\bibfnamefont{N.}~\bibnamefont{Canci}},
  \bibinfo{author}{\bibfnamefont{F.}~\bibnamefont{Carbonara}},
  \bibnamefont{et~al.},
  \bibinfo{title}{Effects of Nitrogen contamination in liquid Argon},
  \bibinfo{journal}{J. Instrum.}
  \textbf{\bibinfo{volume}{5}}, \bibinfo{pages}{P06003}
  (\bibinfo{year}{2010}{\natexlab{b}}).

\bibitem[{\citenamefont{Jones et~al.}(2013)\citenamefont{Jones, Alexander,
  Back, Collin, Conrad, Greene, Katori, Pordes, and Toups}}]{jones2013the}
\bibinfo{author}{\bibfnamefont{B.~J.~P.} \bibnamefont{Jones}},
  \bibinfo{author}{\bibfnamefont{T.}~\bibnamefont{Alexander}},
  \bibinfo{author}{\bibfnamefont{H.~O.} \bibnamefont{Back}},
  \bibinfo{author}{\bibfnamefont{G.}~\bibnamefont{Collin}},
  \bibinfo{author}{\bibfnamefont{J.~M.} \bibnamefont{Conrad}},
  \bibinfo{author}{\bibfnamefont{A.}~\bibnamefont{Greene}},
  \bibinfo{author}{\bibfnamefont{T.}~\bibnamefont{Katori}},
  \bibinfo{author}{\bibfnamefont{S.}~\bibnamefont{Pordes}}, \bibnamefont{and}
  \bibinfo{author}{\bibfnamefont{M.}~\bibnamefont{Toups}},
  \bibinfo{title}{The effects of dissolved methane upon liquid argon scintillation light},
  \bibinfo{journal}{J.
  Instrum.} \textbf{\bibinfo{volume}{8}}, \bibinfo{pages}{P12015}
  (\bibinfo{year}{2013}).

\bibitem[{\citenamefont{Broerman et~al.}(2017)\citenamefont{Broerman, Boulay,
  Cai, Cranshaw, Dering, Florian, Gagnon, Giampa, Gilmour, Hearns
  et~al.}}]{broerman2017application}
\bibinfo{author}{\bibfnamefont{B.}~\bibnamefont{Broerman}},
  \bibinfo{author}{\bibfnamefont{M.}~\bibnamefont{Boulay}},
  \bibinfo{author}{\bibfnamefont{B.}~\bibnamefont{Cai}},
  \bibinfo{author}{\bibfnamefont{D.}~\bibnamefont{Cranshaw}},
  \bibinfo{author}{\bibfnamefont{K.}~\bibnamefont{Dering}},
  \bibinfo{author}{\bibfnamefont{S.}~\bibnamefont{Florian}},
  \bibinfo{author}{\bibfnamefont{R.}~\bibnamefont{Gagnon}},
  \bibinfo{author}{\bibfnamefont{P.}~\bibnamefont{Giampa}},
  \bibinfo{author}{\bibfnamefont{C.}~\bibnamefont{Gilmour}},
  \bibinfo{author}{\bibfnamefont{C.}~\bibnamefont{Hearns}},
  \bibnamefont{et~al.},
  \bibinfo{title}{Application of the {TPB} Wavelength Shifter to the {DEAP}-3600 Spherical Acrylic Vessel Inner Surface},
  \bibinfo{journal}{J. Instrum.}
  \textbf{\bibinfo{volume}{12}}, \bibinfo{pages}{P04017}
  (\bibinfo{year}{2017}).
  
\bibitem[{\citenamefont{Alexander et~al.}(2013)\citenamefont{Alexander, Alton,
  Arisaka, Back, Beltrame, Benziger, Bonfini, Brigatti, Brodsky, Cadonati
  et~al.}}]{Alexander2013light}
\bibinfo{author}{\bibfnamefont{T.}~\bibnamefont{Alexander}},
  \bibinfo{author}{\bibfnamefont{D.}~\bibnamefont{Alton}},
  \bibinfo{author}{\bibfnamefont{K.}~\bibnamefont{Arisaka}},
  \bibinfo{author}{\bibfnamefont{H.}~\bibnamefont{Back}},
  \bibinfo{author}{\bibfnamefont{P.}~\bibnamefont{Beltrame}},
  \bibinfo{author}{\bibfnamefont{J.}~\bibnamefont{Benziger}},
  \bibinfo{author}{\bibfnamefont{G.}~\bibnamefont{Bonfini}},
  \bibinfo{author}{\bibfnamefont{A.}~\bibnamefont{Brigatti}},
  \bibinfo{author}{\bibfnamefont{J.}~\bibnamefont{Brodsky}},
  \bibinfo{author}{\bibfnamefont{L.}~\bibnamefont{Cadonati}},
  \bibnamefont{et~al.},
  \bibinfo{title}{Light yield in DarkSide-10: A prototype two-phase argon TPC for dark matter searches},
  \bibinfo{journal}{Astroparticle Phys.}
  \textbf{\bibinfo{volume}{49}}, \bibinfo{pages}{44 } (\bibinfo{year}{2013}).

\bibitem[{\citenamefont{Rogers}(1974)}]{rogers1974Inelastic}
  \bibinfo{author}{\bibfnamefont{V.~C.} \bibnamefont{Rogers}},
  \bibinfo{title}{Inelastic neutron scattering in $^{19}\mathrm{F}$},
  \bibinfo{journal}{Phys. Rev. C} \textbf{\bibinfo{volume}{9}},
  \bibinfo{pages}{527} (\bibinfo{year}{1974}).

\bibitem[{\citenamefont{Agostinelli et~al.}(2003)\citenamefont{Agostinelli,
  Allison, Amako, Apostolakis, Araujo, Arce, Asai, Axen, Banerjee, Barrand
  et~al.}}]{agostinelli2003geant4}
\bibinfo{author}{\bibfnamefont{S.}~\bibnamefont{Agostinelli}},
  \bibinfo{author}{\bibfnamefont{J.}~\bibnamefont{Allison}},
  \bibinfo{author}{\bibfnamefont{K.}~\bibnamefont{Amako}},
  \bibinfo{author}{\bibfnamefont{J.}~\bibnamefont{Apostolakis}},
  \bibinfo{author}{\bibfnamefont{H.}~\bibnamefont{Araujo}},
  \bibinfo{author}{\bibfnamefont{P.}~\bibnamefont{Arce}},
  \bibinfo{author}{\bibfnamefont{M.}~\bibnamefont{Asai}},
  \bibinfo{author}{\bibfnamefont{D.}~\bibnamefont{Axen}},
  \bibinfo{author}{\bibfnamefont{S.}~\bibnamefont{Banerjee}},
  \bibinfo{author}{\bibfnamefont{G.}~\bibnamefont{Barrand}},
  \bibnamefont{et~al.},
  \bibinfo{title}{{G}eant4—a simulation toolkit},
  \bibinfo{journal}{Nucl. Instrum. Methods Phys. Res., Sect. A}
  \textbf{\bibinfo{volume}{506}}, \bibinfo{pages}{250 } (\bibinfo{year}{2003}),
  ISSN \bibinfo{issn}{0168-9002}.

\bibitem[{\citenamefont{Allison et~al.}(2006)\citenamefont{Allison, Amako,
  Apostolakis, Araujo, Dubois, Asai, Barrand, Capra, Chauvie, Chytracek
  et~al.}}]{allison2006geant4}
\bibinfo{author}{\bibfnamefont{J.}~\bibnamefont{Allison}},
  \bibinfo{author}{\bibfnamefont{K.}~\bibnamefont{Amako}},
  \bibinfo{author}{\bibfnamefont{J.}~\bibnamefont{Apostolakis}},
  \bibinfo{author}{\bibfnamefont{H.}~\bibnamefont{Araujo}},
  \bibinfo{author}{\bibfnamefont{P.~A.} \bibnamefont{Dubois}},
  \bibinfo{author}{\bibfnamefont{M.}~\bibnamefont{Asai}},
  \bibinfo{author}{\bibfnamefont{G.}~\bibnamefont{Barrand}},
  \bibinfo{author}{\bibfnamefont{R.}~\bibnamefont{Capra}},
  \bibinfo{author}{\bibfnamefont{S.}~\bibnamefont{Chauvie}},
  \bibinfo{author}{\bibfnamefont{R.}~\bibnamefont{Chytracek}},
  \bibnamefont{et~al.},
  \bibinfo{title}{Geant4 developments and applications},
  \bibinfo{journal}{IEEE Trans. Nucl. Sci.}
  \textbf{\bibinfo{volume}{53}}, \bibinfo{pages}{270} (\bibinfo{year}{2006}).

\bibitem[{\citenamefont{Saldanha et~al.}(2019)\citenamefont{Saldanha, Back,
  Tsang, Alexander, Elliott, Ferrara, Mace, Overman, and
  Zalavadia}}]{saldanha2019cosmogenic}
\bibinfo{author}{\bibfnamefont{R.}~\bibnamefont{Saldanha}},
  \bibinfo{author}{\bibfnamefont{H.~O.} \bibnamefont{Back}},
  \bibinfo{author}{\bibfnamefont{R.~H.~M.} \bibnamefont{Tsang}},
  \bibinfo{author}{\bibfnamefont{T.}~\bibnamefont{Alexander}},
  \bibinfo{author}{\bibfnamefont{S.~R.} \bibnamefont{Elliott}},
  \bibinfo{author}{\bibfnamefont{S.}~\bibnamefont{Ferrara}},
  \bibinfo{author}{\bibfnamefont{E.}~\bibnamefont{Mace}},
  \bibinfo{author}{\bibfnamefont{C.}~\bibnamefont{Overman}}, \bibnamefont{and}
  \bibinfo{author}{\bibfnamefont{M.}~\bibnamefont{Zalavadia}},
  \bibinfo{title}{Cosmogenic production of $^{39}\mathrm{Ar}$ and $^{37}\mathrm{Ar}$ in argon},
  \bibinfo{journal}{Phys. Rev. C} \textbf{\bibinfo{volume}{100}},
  \bibinfo{pages}{024608} (\bibinfo{year}{2019}).

\bibitem[{\citenamefont{Cleveland et~al.}(1998)\citenamefont{Cleveland, Daily,
  Raymond~Davis, Distel, Lande, Lee, Wildenhain, and
  Ullman}}]{cleveland1998measurement}
\bibinfo{author}{\bibfnamefont{B.~T.} \bibnamefont{Cleveland}},
  \bibinfo{author}{\bibfnamefont{T.}~\bibnamefont{Daily}},
  \bibinfo{author}{\bibfnamefont{J.}~\bibnamefont{Raymond~Davis}},
  \bibinfo{author}{\bibfnamefont{J.~R.} \bibnamefont{Distel}},
  \bibinfo{author}{\bibfnamefont{K.}~\bibnamefont{Lande}},
  \bibinfo{author}{\bibfnamefont{C.~K.} \bibnamefont{Lee}},
  \bibinfo{author}{\bibfnamefont{P.~S.} \bibnamefont{Wildenhain}},
  \bibnamefont{and} \bibinfo{author}{\bibfnamefont{J.}~\bibnamefont{Ullman}},
  \bibinfo{title}{Measurement of the Solar Electron Neutrino Flux with the Homestake Chlorine Detector},
  \bibinfo{journal}{Astrophys. J.} \textbf{\bibinfo{volume}{496}},
  \bibinfo{pages}{505} (\bibinfo{year}{1998}).

\bibitem[{\citenamefont{B{\'e} et~al.}(2013)\citenamefont{B{\'e}, Chist{\'e},
  Dulieu, Mougeot, Chechev, Kondev, Nichols, Huang, and Wang}}]{TabRad_v7}
\bibinfo{author}{\bibfnamefont{M.-M.} \bibnamefont{B{\'e}}},
  \bibinfo{author}{\bibfnamefont{V.}~\bibnamefont{Chist{\'e}}},
  \bibinfo{author}{\bibfnamefont{C.}~\bibnamefont{Dulieu}},
  \bibinfo{author}{\bibfnamefont{X.}~\bibnamefont{Mougeot}},
  \bibinfo{author}{\bibfnamefont{V.}~\bibnamefont{Chechev}},
  \bibinfo{author}{\bibfnamefont{F.}~\bibnamefont{Kondev}},
  \bibinfo{author}{\bibfnamefont{A.}~\bibnamefont{Nichols}},
  \bibinfo{author}{\bibfnamefont{X.}~\bibnamefont{Huang}}, \bibnamefont{and}
  \bibinfo{author}{\bibfnamefont{B.}~\bibnamefont{Wang}},
  \emph{\bibinfo{title}{Table of Radionuclides}}, Vol.~\bibinfo{volume}{7}
  (\bibinfo{publisher}{Bureau
  International des Poids et Mesures}, \bibinfo{address}{S{\`e}vres}, \bibinfo{year}{2013}).

\bibitem[{\citenamefont{Purtschert et~al.}(2017)\citenamefont{Purtschert,
  Kalinowski, Bourgouin, Wieslander, Blanchard, Riedmann, Raghoo,
  Kusmierczyk-Michulec, Gheddou, and adn Yutaka~Tomita}}]{benetti2017Ar}
\bibinfo{author}{\bibfnamefont{R.}~\bibnamefont{Purtschert}},
  \bibinfo{author}{\bibfnamefont{M.}~\bibnamefont{Kalinowski}},
  \bibinfo{author}{\bibfnamefont{P.}~\bibnamefont{Bourgouin}},
  \bibinfo{author}{\bibfnamefont{E.}~\bibnamefont{Wieslander}},
  \bibinfo{author}{\bibfnamefont{X.}~\bibnamefont{Blanchard}},
  \bibinfo{author}{\bibfnamefont{R.}~\bibnamefont{Riedmann}},
  \bibinfo{author}{\bibfnamefont{L.}~\bibnamefont{Raghoo}},
  \bibinfo{author}{\bibfnamefont{J.}~\bibnamefont{Kusmierczyk-Michulec}},
  \bibinfo{author}{\bibfnamefont{A.}~\bibnamefont{Gheddou}}, \bibnamefont{and}
  \bibinfo{author}{\bibfnamefont{C.~S.} \bibnamefont{adn Yutaka~Tomita}}, in
  \emph{\bibinfo{booktitle}{Proceedings of the CTBT Science and Technology 2017 Conference (CTBT,
  Vienna, 2017)}},
  \urlprefix\url{https://ctnw.ctbto.org/ctnw/event/3239/slides/5908D5735D9438FFE05308351BAC687C}.

\bibitem[{\citenamefont{Agnes et~al.}(2018{\natexlab{b}})\citenamefont{Agnes,
  Albuquerque, Alexander, Alton, Araujo, Asner, Ave, Back, Baldin, Batignani
  et~al.}}]{agnes2018Low}
\bibinfo{author}{\bibfnamefont{P.}~\bibnamefont{Agnes}},
  \bibinfo{author}{\bibfnamefont{I.~F.~M.} \bibnamefont{Albuquerque}},
  \bibinfo{author}{\bibfnamefont{T.}~\bibnamefont{Alexander}},
  \bibinfo{author}{\bibfnamefont{A.~K.} \bibnamefont{Alton}},
  \bibinfo{author}{\bibfnamefont{G.~R.} \bibnamefont{Araujo}},
  \bibinfo{author}{\bibfnamefont{D.~M.} \bibnamefont{Asner}},
  \bibinfo{author}{\bibfnamefont{M.}~\bibnamefont{Ave}},
  \bibinfo{author}{\bibfnamefont{H.~O.} \bibnamefont{Back}},
  \bibinfo{author}{\bibfnamefont{B.}~\bibnamefont{Baldin}},
  \bibinfo{author}{\bibfnamefont{G.}~\bibnamefont{Batignani}},
  \bibnamefont{et~al.} (\bibinfo{collaboration}{DarkSide Collaboration}),
  \bibinfo{title}{Low-Mass Dark Matter Search with the DarkSide-50 Experiment},
  \bibinfo{journal}{Phys. Rev. Lett.} \textbf{\bibinfo{volume}{121}},
  \bibinfo{pages}{081307} (\bibinfo{year}{2018}{\natexlab{b}}).

\bibitem[{\citenamefont{Thomas and Imel}(1987)}]{thomas1987recombination}
\bibinfo{author}{\bibfnamefont{J.}~\bibnamefont{Thomas}} \bibnamefont{and}
\bibinfo{author}{\bibfnamefont{D.~A.} \bibnamefont{Imel}},
\bibinfo{title}{Recombination of electron-ion pairs in liquid argon and liquid xenon},
  \bibinfo{journal}{Phys. Rev. A} \textbf{\bibinfo{volume}{36}},
  \bibinfo{pages}{614} (\bibinfo{year}{1987}).

\bibitem[{\citenamefont{Doke et~al.}(1988)\citenamefont{Doke, Crawford,
  Hitachi, Kikuchi, Lindstrom, Masuda, Shibamura, and Takahashi}}]{doke1988let}
\bibinfo{author}{\bibfnamefont{T.}~\bibnamefont{Doke}},
  \bibinfo{author}{\bibfnamefont{H.~J.} \bibnamefont{Crawford}},
  \bibinfo{author}{\bibfnamefont{A.}~\bibnamefont{Hitachi}},
  \bibinfo{author}{\bibfnamefont{J.}~\bibnamefont{Kikuchi}},
  \bibinfo{author}{\bibfnamefont{P.~J.} \bibnamefont{Lindstrom}},
  \bibinfo{author}{\bibfnamefont{K.}~\bibnamefont{Masuda}},
  \bibinfo{author}{\bibfnamefont{E.}~\bibnamefont{Shibamura}},
  \bibnamefont{and}
  \bibinfo{author}{\bibfnamefont{T.}~\bibnamefont{Takahashi}},
  \bibinfo{title}{{Let} dependence of scintillation yields in liquid argon},
  \bibinfo{journal}{Nucl. Instrum. Methods Phys. Res., Sect. A}
  \textbf{\bibinfo{volume}{269}}, \bibinfo{pages}{291 } (\bibinfo{year}{1988}).

\bibitem[{\citenamefont{Berger et~al.}(2010)\citenamefont{Berger, Hubbell,
  Seltzer, Chang, Coursey, Sukumar, Zucker, and Olsen}}]{xcom}
\bibinfo{author}{\bibfnamefont{M.}~\bibnamefont{Berger}},
  \bibinfo{author}{\bibfnamefont{J.}~\bibnamefont{Hubbell}},
  \bibinfo{author}{\bibfnamefont{S.}~\bibnamefont{Seltzer}},
  \bibinfo{author}{\bibfnamefont{J.}~\bibnamefont{Chang}},
  \bibinfo{author}{\bibfnamefont{J.}~\bibnamefont{Coursey}},
  \bibinfo{author}{\bibfnamefont{R.}~\bibnamefont{Sukumar}},
  \bibinfo{author}{\bibfnamefont{D.}~\bibnamefont{Zucker}}, \bibnamefont{and}
  \bibinfo{author}{\bibfnamefont{K.}~\bibnamefont{Olsen}},
  \bibinfo{title}{{XCOM}: Photon cross sections database}
  (\bibinfo{year}{2010}),
  \urlprefix\url{https://www.nist.gov/pml/xcom-photon-cross-sections-database}.

\bibitem[{\citenamefont{Doke et~al.}(2002)\citenamefont{Doke, Hitachi, Kikuchi,
  Masuda, Okada, and Shibamura}}]{Doke2002Absolute}
\bibinfo{author}{\bibfnamefont{T.}~\bibnamefont{Doke}},
  \bibinfo{author}{\bibfnamefont{A.}~\bibnamefont{Hitachi}},
  \bibinfo{author}{\bibfnamefont{J.}~\bibnamefont{Kikuchi}},
  \bibinfo{author}{\bibfnamefont{K.}~\bibnamefont{Masuda}},
  \bibinfo{author}{\bibfnamefont{H.}~\bibnamefont{Okada}}, \bibnamefont{and}
  \bibinfo{author}{\bibfnamefont{E.}~\bibnamefont{Shibamura}},
  \bibinfo{title}{Absolute Scintillation Yields in Liquid Argon and Xenon for Various Particles},
  \bibinfo{journal}{Jpn. J. Appl. Phys.}
  \textbf{\bibinfo{volume}{41}}, \bibinfo{pages}{1538} (\bibinfo{year}{2002}).

\bibitem[{\citenamefont{Lippincott et~al.}(2010)\citenamefont{Lippincott, Cahn,
  Gastler, Kastens, Kearns, McKinsey, and Nikkel}}]{lippincott2010calibration}
\bibinfo{author}{\bibfnamefont{W.~H.} \bibnamefont{Lippincott}},
  \bibinfo{author}{\bibfnamefont{S.~B.} \bibnamefont{Cahn}},
  \bibinfo{author}{\bibfnamefont{D.}~\bibnamefont{Gastler}},
  \bibinfo{author}{\bibfnamefont{L.~W.} \bibnamefont{Kastens}},
  \bibinfo{author}{\bibfnamefont{E.}~\bibnamefont{Kearns}},
  \bibinfo{author}{\bibfnamefont{D.~N.} \bibnamefont{McKinsey}},
  \bibnamefont{and} \bibinfo{author}{\bibfnamefont{J.~A.}
    \bibnamefont{Nikkel}},
  \bibinfo{title}{Calibration of liquid argon and neon detectors with $^{83}\mathrm{Kr}$${}^{m}$},
  \bibinfo{journal}{Phys. Rev. C}
  \textbf{\bibinfo{volume}{81}}, \bibinfo{pages}{045803}
  (\bibinfo{year}{2010}).

\bibitem[{\citenamefont{Agnes et~al.}(2018{\natexlab{c}})\citenamefont{Agnes,
  Dawson, De~Cecco, Fan, Fiorillo, Franco, Galbiati, Giganti, Johnson, Korga
  et~al.}}]{agnes2018measurement}
\bibinfo{author}{\bibfnamefont{P.}~\bibnamefont{Agnes}},
  \bibinfo{author}{\bibfnamefont{J.}~\bibnamefont{Dawson}},
  \bibinfo{author}{\bibfnamefont{S.}~\bibnamefont{De~Cecco}},
  \bibinfo{author}{\bibfnamefont{A.}~\bibnamefont{Fan}},
  \bibinfo{author}{\bibfnamefont{G.}~\bibnamefont{Fiorillo}},
  \bibinfo{author}{\bibfnamefont{D.}~\bibnamefont{Franco}},
  \bibinfo{author}{\bibfnamefont{C.}~\bibnamefont{Galbiati}},
  \bibinfo{author}{\bibfnamefont{C.}~\bibnamefont{Giganti}},
  \bibinfo{author}{\bibfnamefont{T.~N.} \bibnamefont{Johnson}},
  \bibinfo{author}{\bibfnamefont{G.}~\bibnamefont{Korga}}, \bibnamefont{et~al.}
  (\bibinfo{collaboration}{ARIS Collaboration}),
  \bibinfo{title}{Measurement of the liquid argon energy response to nuclear and electronic recoils},
  \bibinfo{journal}{Phys.
  Rev. D} \textbf{\bibinfo{volume}{97}}, \bibinfo{pages}{112005}
  (\bibinfo{year}{2018}{\natexlab{c}}).

\bibitem[{\citenamefont{Xiong et~al.}(2019)\citenamefont{Xiong, Guan, Yang,
  Zhang, Liu, Guo, Wei, Gan, Zhao, and Li}}]{Xiong2019calibration}
\bibinfo{author}{\bibfnamefont{W.-X.} \bibnamefont{Xiong}},
  \bibinfo{author}{\bibfnamefont{M.-Y.} \bibnamefont{Guan}},
  \bibinfo{author}{\bibfnamefont{C.-G.} \bibnamefont{Yang}},
  \bibinfo{author}{\bibfnamefont{P.}~\bibnamefont{Zhang}},
  \bibinfo{author}{\bibfnamefont{J.-C.} \bibnamefont{Liu}},
  \bibinfo{author}{\bibfnamefont{C.}~\bibnamefont{Guo}},
  \bibinfo{author}{\bibfnamefont{Y.-T.} \bibnamefont{Wei}},
  \bibinfo{author}{\bibfnamefont{Y.-Y.} \bibnamefont{Gan}},
  \bibinfo{author}{\bibfnamefont{Q.}~\bibnamefont{Zhao}}, \bibnamefont{and}
  \bibinfo{author}{\bibfnamefont{J.-J.} \bibnamefont{Li}},
  \bibinfo{title}{Calibration of liquid argon detector with $^{83m}Kr$ and $^{22}Na$ in different drift field},
  (\bibinfo{year}{2019}), \eprint{1909.02207}.

\bibitem[{\citenamefont{Miyajima et~al.}(1974)\citenamefont{Miyajima,
  Takahashi, Konno, Hamada, Kubota, Shibamura, and Doke}}]{miyajima1974average}
\bibinfo{author}{\bibfnamefont{M.}~\bibnamefont{Miyajima}},
  \bibinfo{author}{\bibfnamefont{T.}~\bibnamefont{Takahashi}},
  \bibinfo{author}{\bibfnamefont{S.}~\bibnamefont{Konno}},
  \bibinfo{author}{\bibfnamefont{T.}~\bibnamefont{Hamada}},
  \bibinfo{author}{\bibfnamefont{S.}~\bibnamefont{Kubota}},
  \bibinfo{author}{\bibfnamefont{H.}~\bibnamefont{Shibamura}},
  \bibnamefont{and} \bibinfo{author}{\bibfnamefont{T.}~\bibnamefont{Doke}},
  \bibinfo{title}{Average energy expended per ion pair in liquid argon},
  \bibinfo{journal}{Phys. Rev. A} \textbf{\bibinfo{volume}{9}},
  \bibinfo{pages}{1438} (\bibinfo{year}{1974}), \bibinfo{note}{and Phys. Rev. A
  10 (1974) 1452}.

\bibitem[{\citenamefont{Mozumder}(1995)}]{Mozumder1995Free}
  \bibinfo{author}{\bibfnamefont{A.}~\bibnamefont{Mozumder}},
  \bibinfo{title}{Free-ion yield in liquid argon at low-LET},
  \bibinfo{journal}{Chem. Phys. Lett.} \textbf{\bibinfo{volume}{238}},
  \bibinfo{pages}{143 } (\bibinfo{year}{1995}).

\bibitem[{nes()}]{nestweb}
\emph{\bibinfo{title}{{NEST} noble element simulation technique}},
  \urlprefix\url{http://nest.physics.ucdavis.edu}.

\bibitem[{\citenamefont{Aalseth et~al.}(2018)\citenamefont{Aalseth, Acerbi,
  Agnes, Albuquerque, Alexander, Alici, Alton, Antonioli, Arcelli, Ardito
  et~al.}}]{Aalseth2017DarkSide}
\bibinfo{author}{\bibfnamefont{C.~E.} \bibnamefont{Aalseth}},
  \bibinfo{author}{\bibfnamefont{F.}~\bibnamefont{Acerbi}},
  \bibinfo{author}{\bibfnamefont{P.}~\bibnamefont{Agnes}},
  \bibinfo{author}{\bibfnamefont{I.}~\bibnamefont{Albuquerque}},
  \bibinfo{author}{\bibfnamefont{T.}~\bibnamefont{Alexander}},
  \bibinfo{author}{\bibfnamefont{A.}~\bibnamefont{Alici}},
  \bibinfo{author}{\bibfnamefont{A.}~\bibnamefont{Alton}},
  \bibinfo{author}{\bibfnamefont{P.}~\bibnamefont{Antonioli}},
  \bibinfo{author}{\bibfnamefont{S.}~\bibnamefont{Arcelli}},
  \bibinfo{author}{\bibfnamefont{R.}~\bibnamefont{Ardito}},
  \bibnamefont{et~al.},
  \bibinfo{title}{DarkSide-20k: A 20 tonne two-phase LAr TPC for direct dark matter detection at LNGS},
  \bibinfo{journal}{Eur. Phys. J. Plus}
  \textbf{\bibinfo{volume}{133}}, \bibinfo{pages}{131} (\bibinfo{year}{2018}),
  \eprint{1707.08145}.

\bibitem[{\citenamefont{Aramaki et~al.}(2020)\citenamefont{Aramaki, Adrian,
  Karagiorgi, and Odaka}}]{aramaki2020dual}
\bibinfo{author}{\bibfnamefont{T.}~\bibnamefont{Aramaki}},
  \bibinfo{author}{\bibfnamefont{P.~O.~H.} \bibnamefont{Adrian}},
  \bibinfo{author}{\bibfnamefont{G.}~\bibnamefont{Karagiorgi}},
  \bibnamefont{and} \bibinfo{author}{\bibfnamefont{H.}~\bibnamefont{Odaka}},
  \bibinfo{title}{Dual MeV gamma-ray and dark matter observatory - GRAMS Project},
  \bibinfo{journal}{Astropart. Phys.} \textbf{\bibinfo{volume}{114}},
  \bibinfo{pages}{107 } (\bibinfo{year}{2020}).
\end{thebibliography}

\end{document}